\documentclass[journal,twoside,a4paper]{IEEEtran}

\usepackage{amsmath,graphics,graphicx,bm}
\IEEEoverridecommandlockouts
\usepackage{multicol,multirow,bigstrut}
\usepackage{wasysym}
\usepackage{epstopdf,color}
\usepackage{bbm}
\usepackage[utf8]{inputenc} 

\usepackage{amsfonts}

\def\bpsi{\pmb{\psi}_{m|k}}
\def\bvarphi{\pmb{\varphi}}

\def\E{{\rm E}}
\def\sumk{\sum_{j=-k}^k}

\def\T{{\rm T}}

\def\btheta{\pmb{\theta}}

\def\sl{s_{jl}^{m|k}}

\def\btheta{{\bm \theta}}
\def\bvarphi{{\bm \varphi}}

\def\bbeta{{\bm \beta}}

\def\bpsi{{\bm \psi}}

\def\bPhi{{\bm \Phi}}
\def\bPsi{{\bm \Psi}}

\def\bLambda{{\bm \Lambda}}

\def\f{{\rm \bf f}}

\def\h{{\rm \bf h}}

\def\p{{\rm \bf p}}
\def\q{{\rm \bf q}}
\def\R{{\rm \bf R}}
\def\P{{\rm \bf P}}
\def\r{{\rm \bf r}}

\def\ce{{\rm \bf c}}

\def\I{{\rm \bf I}}

\def\F{{\rm \bf F}}
\def\H{{\rm H}}

\def\G{{\rm \bf G}}

\def\Q{{\rm  \bf Q}}

\def\G{{\rm  \bf G}}
\def\g{{\rm  \bf g}}

\title{\LARGE \bf Robust suboptimal local basis function algorithms for identification of nonstationary FIR systems
in impulsive noise environments}

\author{
Maciej Nied\'zwiecki,~\IEEEmembership{Senior Member,~IEEE}, 
Artur Ga\'ncza,~\IEEEmembership{Associate Member,~IEEE},
Wojciech \.Zu\l awi\'nski and Agnieszka Wy\l oma\'nska
\thanks{M. Nied\'zwiecki and A. Ga\'ncza are with 
the Gda\'nsk University of Technology,
Faculty of Electronics, Telecommunications and Informatics,
Department of Signals and Systems, Gda\'nsk, Poland:
{\small  maciekn@eti.pg.edu.pl, artgancz@pg.edu.pl}
\hfill\eject
W. \.Zu\l awi\'nski and A. Wy\l oma\'nska are with the Faculty of Pure and Applied Mathematics, 
Hugo Steinhaus Center, Wroc\l aw University of Science and Technology, 
Wroc\l aw, Poland:
{\small wojciech.zulawinski@pwr.edu.pl, agnieszka.wylomanska@pwr.edu.pl}
\hfill\eject
This work was supported by 
the National Science Center under the agreement 
UMO-2021/40/Q/ST8/00024 (W\.Z, AW).
Computer simulations were carried out at the Academic Computer Centre in Gda\'nsk.}
}

\begin{document}

\maketitle
\thispagestyle{empty}
\pagestyle{empty}

\begin{abstract}
While local basis function (LBF) estimation algorithms, commonly used for identifying/tracking systems with time-varying parameters, demonstrate good performance under the assumption 
of normally distributed measurement noise, the estimation results may significantly deviate from satisfactory when the noise distribution is impulsive in nature, for example, corrupted by outliers. This paper introduces a computationally efficient method to make the LBF estimator robust, enhancing its resistance to impulsive noise. First, the choice of basis functions is optimized based on the knowledge of parameter variation statistics. Then, the parameter tracking algorithm is made robust using the sequential data trimming technique. Finally, it is demonstrated that the proposed algorithm can undergo online tuning through parallel estimation and leave-one-out cross-validation.
\end{abstract}

\vspace{-5mm}

\section{Introduction}

Many nonstationary dynamical systems can be described, or at least well approximated, by linear finite impulse response (FIR) models with
time-varying coefficients.
This is evident, for example, in fading wireless communication channels \cite{tsatsa}--\cite{uwa}, where the FIR model structure arises from multipath propagation (the transmitted signal reaches the receiver along different paths, i.e., with different delays), and the variation in channel parameters is induced by the Doppler effect resulting from the relative movement between the receiver and transmitter and/or surrounding scatterers.

Existing approaches for identifying nonstationary stochastic systems can be broadly categorized into model-free and model-based solutions \cite{book}. In the model-free approach, there is no explicit description of parameter variations. Instead, it is assumed that the unknown system parameters vary at a ``sufficiently slow'' rate, allowing the system to be treated as locally stationary \cite{dahlhaus2012}. Identification is carried out using localized (weighted or windowed) versions of least squares or maximum likelihood estimators, or by employing stochastic gradient algorithms \cite{haykin}--\cite{sayed}.
Nonetheless, such ``unstructured'' identification algorithms, which do not depend on any model of parameter variation, often fall short in terms of achieving adequate estimation accuracy when fast parameter changes are present. In such scenarios, structured estimation methods can be employed, which rely on explicit models of parameter changes, commonly referred to as hypermodels. These hypermodels can be either deterministic or stochastic.
In the deterministic setup, parameter trajectories are approximated using linear combinations of specific functions of time known as basis functions
\cite{rao}--\cite{J58}. Parameter estimates are then obtained by estimating the  approximation coefficients. In the stochastic scenario, a stochastic description of parameter variation is applied, such as the integrated random walk model of a certain order, often referred to as the perturbed polynomial model. Subsequently, after state space embedding, the problem of estimating process parameters can be addressed using algorithms commonly known as Kalman filters/smoothers \cite{norton}--\cite{niedz5}.

In this paper, we advance the deterministic hypermodel-based framework developed in \cite{genSG}
and called the local basis function (LBF) approach. In contrast to \cite{genSG}, where the white additive measurement noise was assumed to be Gaussian, we consider a scenario where the system output is corrupted by additive noise exhibiting impulsive characteristics. We note that such a situation may occur, for example, in some communication  systems \cite{proakis}
and mechanical systems \cite{mechanical}.

In impulsive noise environments the LBF approach, which is sensitive to outliers (occasionally appearing observations particularly large in modulus),
does not yield satisfactory results. 
While there exists an extensive body of statistical literature dedicated to robust regression estimation, the proposed approaches —- ranging from the 
classical idea of M-estimation by Huber \cite{huber} and its many variants based on order statistics (such as L-estimators, R-estimators and S-estimators), to the concepts of least trimmed squares and least median of squares by Rousseeuw \cite{lts} -— 
were primarily designed for batch-type processing of a single frame of data. Due to their computational intensity, often involving iterative searches for solutions, these approaches are not particularly useful when estimation is performed sequentially in a local sliding analysis window, as is the focus of this paper.

Finally, we should acknowledge the existence of a large (and still growing) number of research papers that present recursive adaptive variants of classical robust estimation algorithms
\cite{zou}--\cite{yu}. The common feature of these solutions is that the outlier detection/suppression mechanisms rely solely on the analysis of past input/output data, meaning the corresponding estimation schemes are \textit{causal}. This is not consistent with the LBF framework, which is inherently \textit{non-causal} since the estimation of time-varying system parameters at each instant $t$ is based on both "past" and "future" (relative to $t$) data. Non-causal estimation, which is permissible in some applications, may yield significantly better tracking results than the causal one 
\cite{genSG}. Unlike existing solutions, the robustification technique presented in this paper takes advantage of the benefits of non-causality offered by the LBF scheme.
We propose a computationally affordable method to enhance the robustness of LBF estimators against impulsive measurement noise.

It should be stressed that even though the presented optimization results assume that the measurement noise is a second-order (finite-variance) process, the proposed robustification technique is applicable to a more general class of disturbances, including some types of heavy-tailed (infinite-variance) noise. The presented simulation results confirm that the proposed robust identification algorithms are capable of performing well even in such a more demanding noise environment.

The paper is organized as follows. Section 2 describes the objectives of the paper and their relationship to problems solved in mobile telecommunications. Section 3 presents a brief description of the LBF approach. In Section 4, we demonstrate how to optimize the number and "shape" of basis functions when statistical knowledge of parameter variation is available. Subsequently, in Section 5, we robustify the resulting suboptimal parameter tracking algorithm using the sequential trimming technique. 
Computational complexity of robust algorithms is analyzed in Section 6.
Section 7 introduces a leave-one-out cross-validation-based approach to facilitate the adaptive selection of the trimming level, i.e., the percentage of data points that are removed from each data frame. Section 8 presents simulation results, and Section 9 concludes.

\section{Objectives}

We consider the problem of identification/tracking of a linear nonstationary stochastic system governed by the time-varying FIR model
of the form
\begin{align}
\begin{split}
y(t) &= \sum_{i=1}^n \theta_i^*(t) u(t-i+1) + e(t)\\
&= \btheta^\H(t) \bvarphi(t) + e(t),
\end{split}
\label{1}
\end{align}
where $t= \ldots , -1,0,1, \ldots$ denotes discrete (normalized) time,
$y(t)$ represents the output signal, $u(t)$ represents the input signal, and $e(t)$ represents the white measurement noise. Additionally, $\bvarphi(t)=[u(t),\ldots,u(t-n+1)]^\T$ represents the regression vector, comprising past samples of the observable input signal, and $\btheta(t)=[\theta_1(t),\ldots,\theta_n(t)]^\T$ is the vector
of time-varying system coefficients. All the quantities mentioned above are in the complex domain. 
The symbols $*$ and ${\rm H}$ denote complex conjugate and complex conjugate transpose (Hermitian transpose), respectively. 

One notable practical application that significantly benefits from system identification techniques is self-interference mitigation in full-duplex (FD) underwater acoustic (UWA) communication systems.
In FD UWA systems, both transmit and receive antennas operate simultaneously within the same frequency bandwidth, as described in \cite{shen}--\cite{sparse}. This unique setup nearly doubles the otherwise limited capacity of the acoustic link. In this scenario, the objective is to ``extract'' from $y(t)$ the signal $e(t)$ representing the mixture of the far-end signal and measurement noise. This can be achieved by subtracting from $y(t)$ the self-interference component, which is the first term on the right-hand side of (\ref{1}). This component is associated with reflections of the emitted acoustic wave from surrounding objects, such as the sea surface, sea bottom, fish, vessels, etc. Since the near-end signal $u(t)$, generated by the transmit antenna, is constantly known, the problem essentially boils down to tracking the impulse response coefficients of the self-interference channel.

We will examine a scenario in which the measurement noise $\{e(t)\}$ is a sequence of  complex-valued independent and identically distributed random variables
with zero mean and variance $\sigma_e^2$.
Our goal is to design a robust and computationally affordable estimator for $\btheta(t)$ that minimizes the mean square parameter tracking error (MSE) defined as
$\E\{||\widehat{\btheta}(t) - \btheta(t)||^2\}$.

No specific model of impulsive noise contamination is assumed. 
One commonly used approach is to model $e(t)$ as an $\epsilon$-contaminated white Gaussian noise 
\begin{equation}
e(t) \sim
\left\{
\begin{array}{ccc}
{\mathcal C}{\mathcal N}(0,\sigma_1^2) & {\rm w.p.} & 1-\epsilon \\
{\mathcal C}{\mathcal N}(0,\sigma_2^2) & {\rm w.p.} & \epsilon\ 
\end{array}
\right. \quad \sigma_2^2 \gg \sigma_1^2,
\label{s1a}
\end{equation}
where ${\mathcal C}{\mathcal N}(a,b)$ denotes complex circular Gaussian distribution with mean $a$ and variance $b$, and
$0<\epsilon \ll 1$ denotes a small contamination level.
Alternatively, one can assume that $\{e(t)\}$ is a sequence of random variables with a non-Gaussian leptokurtic distribution. 
It is important to note, however, that the proposed robustification technique, based on sequential data trimming, is not constrained to any specific form of contamination model.

\section{Local basis function approach}

In LBF framework \cite{genSG}, each system parameter is locally approximated through a linear combination of $m$ predefined functions of time, referred to as basis functions. Consequently, the problem of estimating $n$ time-varying system parameters is redefined and solved as an estimation challenge involving $nm$ constant basis expansion coefficients. Unlike the classical basis function approach, which provides interval estimates of the parameter trajectory \cite{functional}, in the LBF framework, estimation is performed independently for each time instant
 $t$, based on the input/output data gathered within the local analysis interval. This leads to a sequence of point estimates, corresponding to different sliding analysis window locations.

In agreement with the local basis function estimation framework, one can assume that the evolution of each system parameter can be locally approximated
by a linear combination of linearly independent functions of time
$f_l(j), j\in I_k=[-k,k], l=1,\ldots,m$. More precisely, it is assumed that at each position of a sliding analysis window
$T_k(t)=[t-k, t+k]$ of width $K=2k+1$, centered at $t$, it holds that
\begin{align}
\begin{split}
\theta_i(t+j) &= \sum_{l=1}^{m} f_l^*(j) b_{il}(t) = \f^\H(j) \bbeta_i(t)\\
j &\in I_k, \quad
i=1,\ldots,n ,
\end{split}
\label{2}
\end{align}
where $\f(j)=[f_1(j), \ldots,f_m(j)]^\T$ and $\bbeta_i(t)= [b_{i1}(t),\ldots,b_{im}(t)]^\T$ denotes the vector of basis expansion coefficients. Since the expansion coefficients
$b_{il}$ may change along with the position of the analysis window, they are written down as functions of $t$.

Typical choices of basis
functions include powers of time
\cite{rao}, \cite{liporace}, \cite{borah} (local Taylor series approximation of the parameter trajectory), harmonic functions  
\cite{functional} (local Fourier approximation)
and wavelets \cite{wav_tsatsanis}--\cite{billings_2002} (wavelet decomposition). 
Without any loss of generality
we assume that basis functions are
orthonormal, i.e.,
\begin{align}
\sumk \f(j)\f^\H(j) = \I_m,
\label{3}
\end{align}
where $\I_m$ denotes the $m\times m$ identity matrix. For any set of basis functions, orthogonalization can be carried out using the Gram-Schmidt procedure.

Let $\bbeta(t) = [b_{11}(t), \ldots,b_{1m}(t), \ldots, b_{n1}(t),\ldots, b_{nm}(t)]^\T$ and 
$\bpsi(t,j) = \bvarphi(t+j)\otimes \f(j)$,
where $\otimes$ denotes Kronecker product of the corresponding matrices/vectors. Combining (\ref{1}) with (\ref{2}) and
using shorthands introduced above, one arrives at the following system description
\begin{align}
y(t+j) = \bbeta^\H(t) \bpsi(t,j) + e(t+j), \ \ j\in I_k.
\label{4}
\end{align}
Based on (\ref{3}), the least squares parameter estimates can be obtained in the form
\begin{align}
\begin{split}
\widehat{\bbeta}_{\rm LBF}(t) &= 
 \operatorname*{arg\,min}_{\bbeta} \sumk  |y(t+j) - \bbeta^{\rm H} \bpsi(t,j)|^2 \\ 
 &=  \P_{\rm LBF}^{-1}(t)\p_{\rm LBF}(t) \\ 
\widehat{\btheta}_{\rm LBF}(t) &= \F_0 \widehat{\bbeta}_{\rm LBF}(t), 
\end{split}
\label{5}
\end{align}
where $\P_{\rm LBF}(t) = \sumk \bpsi(t,j)\bpsi^{\rm H}(t,j)$ is the $mn\times mn$ generalized regression matrix,
$\p_{\rm LBF}(t) = \sumk y^*(t+j)$ $\times\bpsi(t,j)$ is the $mn\times 1$ vector, and $\F_0= \I_n\otimes \f^\H(0)$. 
Since the analysis interval is centered around $t$, i.e., the estimates depend on both ``past'' and ``future'' (relative to $t$)
input/output measurements, the resulting LBF estimators are non-causal. This is a good match for the self-interference
mitigation application since FD UWA systems can be operated both in  block mode and in almost real-time mode (i.e., with a constant decision delay) \cite{uwa1}.
At a qualitative level, the LBF approach can be regarded as an extension, specifically tailored for system identification, of the signal smoothing technique known as Savitzky-Golay (SG) filtering \cite{schafer}.

\section{Optimization}

\subsection{Parameter tracking}
\noindent{To} leverage the known properties of LBF estimators, we need to make several technical assumptions:
\begin{enumerate}
\item[({A1})]
$\{u(t)\}$ is a zero-mean wide-sense stationary Gaussian sequence with an autocorrelation function that decays exponentially and a variance of $\sigma_u^2$.
\item[({A2})]
The sequence $\{e(t)\}$, independent of $\{u(t)\}$, consists of zero-mean independent and identically distributed random variables with a variance of $\sigma_e^2$.
\item[({A3})]
The sequence $\{\btheta(t)\}$ is independent of  $\{u(t)\}$ and  $\{e(t)\}$.
\end{enumerate}
As shown in \cite{genSG}, under (A1)-(A3) the following approximations hold true
\begin{align}
\bar{\btheta}_{\rm LBF}(t) = \E\left[\widehat{\btheta}_{\rm LBF}(t)\right] \cong \sumk h(j) \btheta(t+j)
\label{o1}
\end{align}
\begin{align}
&\hspace{20mm}{\rm cov} \left[\widehat{\btheta}_{\rm LBF}(t)\right] \nonumber\\
&=\E\left[\left(\widehat{\btheta}_{\rm LBF}(t) - \bar{\btheta}_{\rm LBF}(t) \right) 
\left(\widehat{\btheta}_{\rm LBF}(t) - \bar{\btheta}_{\rm LBF}(t) \right)^\H \right]\nonumber\\
&\hspace{20mm}\cong \sigma_e^2 \bPhi^{-1} \sumk |h(j)|^2,
\label{o2}
\end{align}
where $\{h(j), j\in I_k\}$ denotes the basis-dependent impulse response of the linear smoothing filter
\begin{align}
h(j) = \f^\H(0) \f(j),
\label{s6}
\end{align}
$\bPhi={\rm cov}[\bvarphi(t)]$, and averaging is carried over $\{u(t)\}$ and $\{e(t)\}$.
The accuracy of the second approximation increases with the number of included basis functions.

\vskip 2mm
\noindent{\bf Remark 1}
\vskip 1mm
\noindent{The} assumption of Gaussianity of the input signal imposed by assumption (A1) is not a critical requirement. To arrive at (\ref{o1})-(\ref{o2}), 
one can equally well assume that \cite{thesis}:
\vspace{2mm}
\begin{enumerate}
\item[({${\rm A1}^\star$})]
$\{u(t)\}$ is a zero-mean wide-sense stationary $l$-dependent process
 with a variance of $\sigma_u^2$ and finite fourth-order moments.
\end{enumerate}
\vspace{-4mm}
\hfill\rule{2mm}{2mm}

\vspace{2mm} 
\noindent{The} relationships (\ref{o1})-(\ref{o2}) pave the way for optimizing parameter tracking when prior knowledge about the variation of 
system parameters is available. As an example, we will consider the following practically significant scenario:
\begin{enumerate}
\item[({A4})]
$\{\theta_i(t)\}$, $i=1,\ldots,n$ are zero-mean wide-sense stationary processes with known autocorrelation functions
$\E[\theta_i(t+\tau) \theta_i^*(t)]= \sigma_{\theta_i}^2 \rho_\theta(\tau)$, $i=1,\ldots,n$.
\end{enumerate}
Assumption (A4) is fulfilled in many mobile communication systems. For radio communication channels operating under stationary scattering and Rayleigh fading, 
the Jakes' model is often employed, with a Doppler spectrum of the form \cite{jakes}
\begin{align}
S_{\theta_i}(\omega) = \left\{
\begin{array}{ccc}
\frac{2 \sigma_{\theta_i}^2}{\omega_d} \sqrt{\frac{1}{1-\left(\frac{\omega}{\omega_d}\right)^2}} & {\rm for} & |\omega|\le \omega_d\\
0 & {\rm for} & |\omega|> \omega_d
\end{array}
\right.
\label{o3}
\end{align}
where $\omega_d$ is inversely proportional to the vehicle speed. In this case the normalized autocorrelation function of channel parameters
has the form $\rho_\theta(\tau)=J_0(\omega_d \tau)$ where
$J_0(\cdot)$ denotes the zero order Bessel function.

On the other hand, for underwater acoustic channels, the flat Doppler spectrum model is frequently adopted \cite{uwa1}
\begin{align}
S_{\theta_i}(\omega) = \left\{
\begin{array}{ccc}
\frac{\pi \sigma_{\theta_i}^2}{\omega_0} & {\rm for} & |\omega|\le \omega_0\\
0 & {\rm for} & |\omega|> \omega_0
\end{array}
\right.
\label{o4}
\end{align}
where $\omega_0$ depends on environmental factors. The resulting normalized autocorrelation  function is given by
$\rho_\theta(\tau)= {\rm sinc} (\omega_0 \tau)$, where ${\rm sinc}(\cdot)$ denotes the sine cardinal function.

In both of the cases discussed above, channel parameters can be seen as mutually independent processes, 
an assumption that is not required for our further developments.

Following \cite{KL}, \cite{tsp24}, we  choose as basis functions the dominant eigenvectors of the $K\times K$
correlation matrix $\R_\theta$,  $[\R_\theta]_{ij}= \rho_\theta(j-i), i,j=1,\ldots, K$, namely the ones that correspond to the $m$
largest eigenvalues of $\R_\theta$. 
Denote by $\q_1,\ldots, \q_K$ the eigenvectors of $\R_\theta$ arranged in the order of decreasing eigenvalues
$\lambda_1 \ge \ldots \ge\lambda_K >0$. The basis sequences will be chosen according to
\begin{align}
[f_i(k), \ldots, f_i(-k)]^\T = \q_i, \quad i=1,\ldots,m.
\label{o5a}
\end{align}
We demonstrate that such a choice, motivated by the well-known Karhunen-Lo\`eve expansion theorem \cite{haykin}, enables the derivation of a simple rule for selecting $m$ to minimize the mean squared parameter tracking error
\begin{align}
{\rm MSE}_{\rm LBF}(m) &= \E\left[|| \btheta(t) - \bar{\btheta}_{\rm LBF}(t)||^2 \right]\nonumber\\
+ {\rm tr} \left\{{\rm cov}[\widehat{\btheta}_{\rm LBF}(t)]  \right\} &= B_{\rm LBF}(m) + V_{\rm LBF}(m),
\label{o5}
\end{align}
where the first term on the right-hand side of (\ref{o5}) denotes the bias component of MSE, and the second term is its variance component.
The expectation in (\ref{o5}) is taken over all realizations of the parameter trajectory $\{\btheta(t)\}$.

Since the matrix $\R_\theta$ is Hermitian, it holds that $f_i(-j)=f_i^*(j), j=0,\ldots,k$, which implies that the quantities
$f_i(0), i=1,\ldots,K$, are real valued.
Note also, that to guarantee system identifiability, i.e., nonsingularity of the $mn\times mn$ generalized regression matrix $\P_{\rm LBF}(t)$,
the number of basis functions must be restricted to (at most) $M=K/n$.

It can be shown that (see Appendix 1)
\begin{align}
 B_{\rm LBF}(m) \cong \sigma_\theta^2\left[1 - \sum_{i=1}^m \lambda_i f_i^2(0) \right]
\label{o6a}
\end{align}
\begin{align}
V_{\rm LBF}(m) \cong \sigma_e^2 {\rm tr} \left\{ \bPhi^{-1} \right\} \sum_{i=1}^m  f_i^2(0),
\label{o7a}
\end{align}
where $\sigma_\theta^2= \E[||\btheta(t)||^2] = \sum_{i=1}^n \sigma_{\theta_i}^2$.

Note that the bias component of MSE decreases while its variance component increases with growing $m$, 
which reflects the well-known bias-variance tradeoff typical of identifying time-varying systems.
Minimization of ${\rm MSE}_{\rm LBF}(m)$ with respect to $m$ is straightforward leading to
\begin{align}
m_{\rm opt} = \operatorname*{arg\,max}_{1\le m < M} \left\{ m\ {\rm s.t.}\ \lambda_m > \frac{\sigma_e^2 {\rm tr}\{\bPhi^{-1}\}}{\sigma_\theta^2}  \right\}.
\label{o8}
\end{align}

When no prior information about the system/signal statistics is available, necessitating the selection of basis functions solely based on universal 
approximation principles, optimizing the tracking/smoothing performance of the estimation algorithm becomes a challenging problem with two degrees of freedom. 
As demonstrated in \cite{genSG}, when the number of basis functions $m$ is held constant and the analysis window size $K$ is increased, the variance component decreases while the bias component increases. Similarly, for a fixed window size, increasing the number of basis functions allows for a reduction in the bias component, but concurrently increases the variance component. Consequently, it is evident that the values of $m$ and $K$ should be chosen to strike a balance between the bias and variance components of the MSE. However, an analytical solution to 
this joint optimization problem is not known. The practical solution is based on parallel estimation – several algorithms corresponding to different choices of $K$ and $m$ are run concurrently, and at each time instant, the best-local estimates are selected according to some statistical criteria, such as cross-validation or Akaike's final prediction error (FPE) statistic \cite{genSG}.

In the case examined in this paper, the scenario differs. With the number of basis functions being automatically adjusted for a chosen window size $K$, the focus shifts to optimizing the value of $K$. This task, however, is straightforward since, under assumption (A4), the mean squared estimation error uniformly decreases with increasing $K$. Nevertheless, this does not 
imply that $K$ should be arbitrarily large. Apart from the increased computational load, extending the width of the analysis window beyond a certain value may yield only marginal performance improvements due to the saturation effect.

\section{Robustification}

\subsection{Basic algorithm}
While LBF algorithms perform well when the measurement noise is Gaussian, the results of tracking/smoothing may be far from satisfactory when the distribution of 
${e(t)}$  is of impulsive nature.

To enhance the robustness of LBF estimators, making them less susceptible to impulsive noise, we employ a method known as trimming. Trimming involves removing from the data set used for estimation those samples that correspond to excessively large modeling errors. This technique is well-established in signal and image processing, see e.g. \cite{lts}, \cite{bovik}.
Let 
 \begin{align}
 \widetilde{K}= {\rm int}[\gamma K] \ge mn,
 \label{9b} 
 \end{align}
 where $0<\gamma <1$ and ${\rm int}[x]$ is the greatest integer contained in $x$, denote the number of data points
 that will not be removed from the analyzed data frame. 
The number of rejected samples, equal to $K-\widetilde{K}$, will be denoted by $\delta$. Furthermore, denote by $\Omega(t)\subset \Omega_0 = \{-k, \ldots,k\}$,
${\rm card}[\Omega(t)]= \widetilde{K}$ 
the set indicating the position, within the interval $[-k,k]$, of samples that are used for estimation purposes
in the analysis interval $T_k(t)$. Similarly, let $\bar{\Omega}(t)= \Omega_0- \Omega(t)$, ${\rm card}[\bar{\Omega}(t)]= K-\widetilde{K}$, denote the set indicating positions of 
$\delta$ samples that are trimmed away.

The most sophisticated form of trimming is the least trimmed squares (LTS) approach \cite{lts}, which entails identifying 
the subset of $\widetilde{K}$ regressors, out of all $K$ regressors, that yields the lowest sum of squared residuals 
\begin{align}
\begin{split}
\widehat{\bbeta}_{\rm LTS}(t) &= 
 \operatorname*{arg\,min}_{\bbeta, \Omega} \sum_{j\in \Omega}  |y(t+j) - \bbeta^{\rm H} \bpsi(t,j)|^2 \\ 
\widehat{\btheta}_{\rm LTS}(t) &= \F_0 \widehat{\bbeta}_{\rm LTS}(t), 
\end{split}
\label{5a}
\end{align}
where ${\rm card}[\Omega]= \widetilde{K}$.

Since, like all subset regression approaches, the LTS framework is computationally very intensive, particularly for large data sets, 
it is not well-suited for sequential processing \cite{lts1}. For this reason, rather than employing LTS estimators, we will utilize sequentially 
trimmed LBF estimators as described below.

Denote by $\widehat{\bbeta}_{\rm R}(t-1)$ the robust estimate
of $\bbeta$ obtained using the data from the analysis frame $T_k(t-1)$ and denote by
\begin{align}
\begin{split}
\varepsilon(t-1+j|t-1) &= y(t-1+j) - \widehat{\bbeta}_{\rm R}^\H(t-1)\bpsi(t-1,j)\\
 j &=-k+1, \ldots, k
\end{split}
\label{7}
\end{align}
the resulting residual errors. Additionally
denote by
\begin{align*}
\varepsilon(t+k|t-1) &= y(t+k) - \widehat{\bbeta}_{\rm R}^\H(t-1)\bpsi(t,k)
\end{align*}
the modeling error anticipated at the instant $t+k$. 
The decision, made sequentially for consecutive values of $t$, regarding which samples can be accepted in the frame $T_k(t)$, can be based on
rejection of samples corresponding to $\delta$ largest modeling errors observed in the preceding frame.
Denote by
\begin{align}
\{ \widetilde{\varepsilon}(t-k|t-1), \ldots, \widetilde{\varepsilon}(t+k|t-1)\}
\label{13}
\end{align}
the sequence obtained by rearranging the sequence
\begin{align}
\{ {\varepsilon}(t-k|t-1), \ldots, {\varepsilon}(t+k|t-1)\}
\label{14}
\end{align}
in the order of non-decreasing modulus
\[
 |\widetilde{\varepsilon}(t-k|t-1)| \le \ldots \le |\widetilde{\varepsilon}(t+k|t-1)| .
\]
Finally, let $o(\cdot) : \{-k,\ldots,k\} \longmapsto \{-k,\ldots,k\}$ be the mapping function that shows the position of the error $\varepsilon(t+j|t-1)$
in the ordered sequence of errors
\[
\varepsilon(t+j|t-1) = \widetilde{\varepsilon}(t+o(j)|t-1).
\]
Using this notation, the set $\Omega(t)$ can be defined as follows
\begin{align}
\Omega(t) = \{ j\in I_k \ {\rm s.t.\ } o(j) \le k-\delta\}.
\label{15}
\end{align}

Adopting this strategy leads to the following estimation formula, which is further referred to as the sequentially trimmed LBF estimator
\begin{align}
\begin{split}
\widehat{\bbeta}_{\rm R}(t) &= 
 \operatorname*{arg\,min}_{\bbeta} \sum_{j\in\Omega(t)}  |y(t+j) - \bbeta^{\rm H} \bpsi(t,j)|^2 \\ 
 &=  \P_{\rm R}^{-1}(t)\p_{\rm R}(t) \\ 
\widehat{\btheta}_{\rm R}(t) &= \F_0 \widehat{\bbeta}_{\rm R}(t) ,
\end{split}
\label{11}
\end{align}
where
\begin{align}
\begin{split}
\P_{\rm R}(t) &= \sum_{j\in\Omega(t)} \bpsi(t,j)\bpsi^\H(t,j)\\
&= \P_{\rm LBF}(t) - \sum_{j\in \bar{\Omega}(t)} \bpsi(t,j)\bpsi^\H(t,j)\\
\p_{\rm R}(t) &= \sum_{j\in\Omega(t)} y^*(t+j)\bpsi(t,j)\\
&= \p_{\rm LBF}(t) - \sum_{j\in \bar{\Omega}(t)} y^*(t+j)\bpsi(t,j).
\end{split}
\label{12}
\end{align}

For an initial estimate of $\bbeta$ at time $t_0$, a viable approach is to utilize the LTS estimate by setting $\widehat{\bbeta}_{\rm R}(t_0) = \widehat{\bbeta}_{\rm LTS}(t_0)$, or opt for computationally less intensive "batch" robust estimates, such as the least absolute deviation (LAD) estimate as described below.
\noindent{The} estimation formula (\ref{4}) is derived from the minimization of the $\ell_2$ norm 
of residuals. One possible way to modify LBF estimators is by replacing the least squares estimates in (\ref{4}) with their least absolute deviations counterparts
\begin{align}
\begin{split}
\widehat{\bbeta}_{\rm LAD}(t) &= 
 \operatorname*{arg\,min}_{\bbeta} \sumk  |y(t+j) - \bbeta^{\rm H} \bpsi(t,j)|\\
\widehat{\btheta}_{\rm LAD}(t) &= \F_0 \widehat{\bbeta}_{\rm LAD}(t) .
\end{split}
\label{a6}
\end{align}
The LAD estimator is renowned for its robustness, displaying reduced sensitivity to outliers in the data compared to the least squares estimator.
Unfortunately, when the $\ell_2$ norm is replaced with the $\ell_1$ norm, a numerical search for the solution becomes necessary.
For this reason, the estimate (\ref{a6}) is further used only as a starting point for recursive estimation and as a reference for 
comparative purposes.

\subsection{Adaptive version}
If the quantities $\sigma_e^2$ and $\sigma_\theta^2$ are not known, or change slowly over time, they can be replaced in (\ref{o8}) with their local
estimates.

Observe that
\begin{align}
\begin{split}
&\varepsilon(t+j|t)= \left[\bbeta(t) - \widehat{\bbeta}_{\rm R}(t) \right]^\H\bpsi(t,j) + e(t+j)\\
&= \left[ \btheta(t+j) - \widehat{\btheta}_{\rm R}(t+j|t) \right]^\H \bvarphi(t+j) +e(t+j)\\
&\hspace{2cm} j \in I_k,
\end{split}
\label{8}
\end{align}
where $\{\widehat{\btheta}_{\rm R}(t+j|t)=\F(j)\widehat{\bbeta}_{\rm R}(t), \F(j)=\I_n\otimes \f^\H(j), j\in I_k\}$ denotes
the estimated parameter trajectory in the entire 
interval $[t-k,t+k]$. Therefore, when parameter tracking is satisfactory so that the
first term on the right-hand side of (\ref{8}) can be neglected, one obtains $\varepsilon(t+j|t)\cong e(t+j)$.

Based on this observation, the local noise variance estimate can be obtained in the form
\begin{align}
\widehat{\sigma}_e^2(t) = \frac{1}{\widetilde{K}} \sum_{j\in \Omega(t)} |\varepsilon(t+j|t)|^2.
\label{100}
\end{align}
We note that, when $\varepsilon(t+j|t)\equiv e(t+j)$, (\ref{100}) is a consistent estimator of the so-called quantile conditional variance of $e(t)$, which is well-defined and finite for
a wide range of distributions, including all heavy-tailed (infinite-variance) distributions that belong to the $\alpha$-stable family \cite{quantile}.

Similarly, the local variance of parameter  changes can be estimated using the following formula
\begin{align}
\widehat{\sigma}_\theta^2(t) = \frac{1}{K} \sumk ||\widehat{\btheta}_{\rm R}(t+j|t) - \bar{\btheta}_{\rm R}(t)||^2
\label{101}
\end{align}
where
\begin{align}
\bar{\btheta}_{\rm R}(t) = \frac{1}{K} \sumk \widehat{\btheta}_{\rm R}(t+j|t).
\label{102}
\end{align}
It can be shown (see Appendix 2) that the estimate (\ref{101}) can be easily computed in terms
of $\widehat{\bbeta}_{\rm R}(t)$
\begin{align}
\widehat{\sigma}_\theta^2(t)  = \frac{1}{K} ||\widehat{\bbeta}_{\rm R}(t)||^2 - ||\G \widehat{\bbeta}_{\rm R}(t)||^2,
\label{103}
\end{align}
where $\G= \I_n \otimes \g^\H$, $\g= (1/K) \sumk \f(j)$.

Combining (\ref{o8}) with (\ref{100}) and (\ref{101}), one arrives at
the following adaptive rule for selection of the number of basis functions
\begin{align}
\widehat{m}(t) = \operatorname*{arg\,max}_{1\le m < \widetilde{M}} \left\{ m\ {\rm s.t.}\ \lambda_m > 
\frac{\widehat{\sigma}_e^2(t-1) {\rm tr}\{{\bPhi}^{-1}\}}{\widehat{\sigma}_\theta^2(t-1)}  \right\},
\label{o10}
\end{align}
where $\widetilde{M}=\widetilde{K}/n$.

Finally, if the covariance matrix $\bPhi$ is slowly time-varying (in communication applications it is usually known and time-invariant), 
it can be replaced in (\ref{o10}) 
with its exponentially weighted estimate
\begin{align}
\widehat{\bPhi}(t) = \eta_0 \widehat{\bPhi}(t-1) + (1-\eta_0) \bvarphi(t)\bvarphi^\H(t),
\label{104}
\end{align}
where $\eta_0 \in (0,1)$ denotes the forgetting constant.

Note that the rule (\ref{o10}) can also be utilized to create an adaptive version of the LAD algorithm, 
which is further used in the simulation section for comparative purposes.

\subsection{Extensions}
The robust tracking/smoothing solution proposed in this section generalizes straightforwardly to the problem of system identification in the presence of 
missing samples. In this scenario, the set $\Omega(t)$ is known \textit{a priori} for each position of the analysis window. The same approach can also be applied 
when some samples, with known locations, are missing, and others, with unknown locations, are corrupted by outliers.

\section{Computational complexity}

\subsection{Matrix eigendecomposition}

The numerically reliable method for eigendecomposition of the correlation matrix $\mathbf{R}_\theta$ involves transforming it into tridiagonal form. The eigenvalues, along with their corresponding eigenvectors, can be computed sequentially from largest to smallest using the Lanczos algorithm \cite{lanczos1, lanczos2}. The associated computational complexity is approximately ${\mathcal O}(mK^2)$, measured in terms of complex MAC (multiply and accumulate) operations.
If the Toeplitz structure of $\mathbf{R}_\theta$ is taken advantage of, this cost can be further reduced to ${\mathcal O}(K^2)$ \cite{trench}.

\subsection{Parameter tracking}

In the regular LBF algorithm, the computational load per time update consists of the following components: $Km^2n^2$ (for computing $\P_{\rm LBF}$),
$Kmn$ (for computing $\p_{\rm LBF}$), ${\mathcal O}(m^3n^3)$ (for computing $\widehat{\bbeta}_{\rm LBF}$), and
$mn$ (for computing $\widehat{\btheta}_{\rm LBF}$).

When implementing the robust LBF algorithm, the computational load increases by: $2Kmn$ (for computing residual errors) and
${\mathcal O}(K\log \delta)$ (for localizing the $\delta$ largest residuals). 
The cost of evaluating $\P_{\rm R}$, $\p_{\rm R}$, $\widehat{\bbeta}_{\rm R}$ and $\widehat{\btheta}_{\rm R}$
is slightly lower than the cost of evaluating $\P_{\rm LBF}$, $\p_{\rm LBF}$, $\widehat{\bbeta}_{\rm LBF}$ and $\widehat{\btheta}_{\rm LBF}$,
respectively.

\section{Selection of trimming level}

The mean square parameter tracking error depends on the choice of the trimming level $\delta$. Consider a scenario where the width $K$ of the local analysis window 
is fixed, and the goal is to determine the most suitable value for $\delta$. When $\delta$ is too small, certain large outliers may persist, causing substantial errors in 
parameter estimation. Conversely, opting for an excessively large $\delta$ also poses challenges, as it reduces the number of samples used for parameter estimation. Hence, it is 
essential to seek a trade-off value for $\delta$ to optimize tracking performance.
Due to the analytical complexity of the problem, direct optimization of $\delta$ does not seem possible. The solution proposed below is based on parallel estimation and involves choosing $\delta$ from the finite set of pre-defined values $\{\delta_1, \ldots, \delta_p\}$, respectively.

As an instantaneous measure of the estimation accuracy at time instant $t$, one can utilize the leave-one-out output interpolation error, 
often referred to as deleted residual. Suppose that $0 \in \Omega(t)$, indicating that the system output is classified as outlier-free. The deleted residual is defined as follows
\begin{align*}
\varepsilon^\circ(t|t) &= y(t) - [\widehat{\btheta}^\circ_{\rm R}(t)]^\H \bvarphi(t)
= y(t) - [\widehat{\bbeta}^\circ_{\rm R}(t)]^\H \bpsi(t,0),
\end{align*}
where $\widehat{\btheta}^\circ_{\rm R}(t)$ and $\widehat{\bbeta}^\circ_{\rm R}(t)$ denote the ``holey'' estimates of $\btheta(t)$ and $\bbeta(t)$, respectively,
obtained after excluding the observation collected at the time instant $t$ from the local dataset
\begin{align}
\begin{split}
\widehat{\bbeta}_{\rm R}^\circ(t) &= 
 \operatorname*{arg\,min}_{\bbeta} \sum_{j\in\Omega^\circ(t)}  |y(t+j) - \bbeta^{\rm H} \bpsi(t,j)|^2 \\ 
 &=  [\P_{\rm R}^\circ(t)]^{-1}\p_{\rm R}^\circ(t) \\ 
\widehat{\btheta}_{\rm R}^\circ(t) &= \F_0 \widehat{\bbeta}_{\rm R}^\circ(t), 
\end{split}
\label{23}
\end{align}
where $\Omega^\circ(t)=\Omega(t) -\{0\}$ and
\begin{align}
\begin{split}
\P_{\rm R}^\circ(t) &= \P_{\rm R}(t) -  \bpsi(t,0)\bpsi^\H(t,0)\\
\p_{\rm R}^\circ(t) &= \p_{\rm R}(t) - y^*(t)\bpsi(t,0).
\end{split}
\label{24}
\end{align}
Deleted residuals can be expressed in a computationally more convenient form. Let
\begin{align}
c(t) = \bpsi^\H(t,0) \P_{\rm R}^{-1}(t) \bpsi(t,0).
\label{25}
\end{align}
Using the matrix inversion lemma \cite{haykin}, one arrives at
\begin{align}
[\P_{\rm R}^\circ(t)]^{-1} = \P_{\rm R}^{-1}(t)
+ \frac{ \P_{\rm R}^{-1}(t)\bpsi(t,0)\bpsi^\H(t,0) \P_{\rm R}^{-1}(t)}{1-c(t)}.
\label{26}
\end{align}
Combining (\ref{26}) with (\ref{24}) and (\ref{25}), one obtains, after elementary calculations
\begin{align*}
\widehat{\bbeta}_{\rm R}^\circ(t)  = \bbeta_{\rm R}(t) - [y^*(t) - \bpsi^\H(t,0)\widehat{\bbeta}_{\rm R}(t)]
\frac{\P_{\rm R}^{-1}(t) \bpsi(t,0)}{1-c(t)}
\end{align*}
which finally leads to
\begin{align}
\varepsilon^\circ(t|t) = \frac{\varepsilon(t|t)}{1-c(t)}\; ,
\label{27}
\end{align}
where
\begin{align*}
\varepsilon(t|t) = y(t) - \widehat{\bbeta}_{\rm R}^\H(t) \bpsi(t,0) 
=y(t) - \widehat{\btheta}_{\rm R}^\H(t) \bvarphi(t)
\end{align*}
denotes the residual error. According to (\ref{27}), deleted residuals 
can be evaluated in a straightforward manner 
without the need to implement the holey estimation scheme.

Deleted residuals can be used for tuning robust LBF algorithms. The proposed solution is based on parallel estimation and cross-validation.
Consider $p$ estimation algorithms, each equipped with a different  trimming coefficient $\delta_1 < \delta_2 < ... < \delta_p$, 
working concurrently and producing the estimates $\widehat{\bbeta}_{{\rm R}|i}(t)= \P_{{\rm R}|i}^{-1}(t)\p_{{\rm R}|i}(t)$, 
$\widehat{\btheta}_{{\rm R}|i}(t)= \F_0\widehat{\bbeta}_{{\rm R}|i}(t)$ for $i=1,...,p$.
Denote by $\varepsilon_i(t|t)$ and $\varepsilon_i^\circ(t|t)$ the associated residuals and deleted residuals, respectively.

In the absence of outliers, as a local measure that can be used to select the best-fitting solution, one can utilize 
the sum of the squared $L$ most recent deleted residuals, leading to
\vspace{-1mm}
\begin{align}
i_*(t) =  \operatorname*{arg\,min}_{1 \le i\le p} {\mathcal E}_i(t),
\label{28}
\end{align}
\vspace{-5mm}
\begin{align}
\begin{split}
{\mathcal E}_i(t) &= \sum_{l=0}^{L-1} |\varepsilon_i^\circ(t-l|t-l)|^2 = {\mathcal E}_i(t-1) \\
&- |\varepsilon_i^\circ(t-L|t-L)|^2
+ |\varepsilon_i^\circ(t|t)|^2.
\end{split}
\label{29}
\end{align}
Such a selection procedure is usually referred to as leave-one-out cross-validation.
The value of $L$ should be sufficiently large to ensure statistical reliability of the selection results,
but not too large to maintain the ability of the selection procedure to keep pace with the possible changes in system and noise characteristics.
The typical choice for $L$ is $L\in [30,50]$.

In the presence of outliers, the cross-validation procedure must be modified to exclude data samples corrupted by excessively large measurement errors from the decision process. To achieve this, we will utilize only those deleted residuals, hereafter referred to as agreed deleted residuals, which are considered outlier-free by all competing estimation algorithms. In other words, these residuals that satisfy the condition
\vskip -3mm
\[
0 \in \widetilde{\Omega}(t) = \bigcap_{i=1}^p \Omega_i(t).
\]
\vskip -1mm
For this reason, the selection decisions will be based on the minimization of the $L$ most recent agreed residuals instead of the 
minimization of just the $L$ most recent ones, as in (\ref{29}). 

The recursive algorithm for updating the modified statistic ${\mathcal E}_i(t)$ can be easily constructed with the aid of a circular register
$\xi_i(0),...,\xi_i(L-1)$ which keeps track of, chronologically ordered, $L$ most recent agreed deleted residuals. Denote by $l\in[0,L-1]$ the current position within this register
of the ``oldest'' deleted residual, successively replaced by the ``newest'' one. The pseudo-code of the updating algorithm is as follows
\begin{align}
\begin{split}
&{\rm \bf for} \ t=t_0+1, t_0+2, ... \ {\rm  \bf do}\\
&\hspace{3mm} {\rm  \bf if} \ 0\in \widetilde{\Omega}(t)\\
&\hspace{6mm} {\mathcal E}_i(t) = {\mathcal E}_i(t-1) - |\xi_i(l)|^2 + |\varepsilon_i^\circ(t|t)|^2\\
&\hspace{6mm} \xi_i(l) = \varepsilon_i^\circ(t|t)\\
&\hspace{6mm} l:=(l+1)\;{\rm mod}\; L\\
&\hspace{3mm} {\rm  \bf else}\\
&\hspace{6mm} {\mathcal E}_i(t) = {\mathcal E}_i(t-1)\\
&\hspace{3mm} {\rm  \bf end\ if}\\
& {\rm  \bf end}
\end{split}
\label{29a}
\end{align}
where $(l+1){\rm mod} L$ denotes the remainder after dividing $l+1$ by $L$ (modulo operation).
Assuming that at the instant $t_0$, the oldest deleted residual is stored in $\xi_i(0)$ and the newest one is 
stored in $\xi_i(L-1)$, the initial value of $l$ should be set to 0.

When the best-local value of $i$ is selected using (\ref{28}) and (\ref{29a}), the estimate of $\btheta(t)$ 
takes the form
\begin{align}
\widehat{\btheta}_{\rm A}(t) = \widehat{\btheta}_{{\rm R}|i_*(t)}(t).
 \label{31}
\end{align}
The unquestionable advantage of cross-validation is that it provides sensible results regardless of whether the assumptions
underlying the compared estimation methods are met
{or} not. 

\subsection*{Computational optimizations}

Let $\Omega_i(t)$ denote the set indicating the positions of samples used for estimation in the interval $T_k(t)$ when the trimming level is set to $\delta_i$. 
Additionally, let $\bar{\Omega}_i(t) = \Omega_0 - \Omega_i(t)$ represent the set indicating the positions of $\delta_i$ samples that are trimmed away.

Define 
\begin{align*}
\Delta_i(t) &= \bar{\Omega}_i(t) - \bar{\Omega}_{i-1}(t) = \{j_1(t),\ldots,j_{r_i}(t)\}\\  
i&=2,\ldots,p
\end{align*} 
where $r_i = {\rm card}[\Delta_i(t)] = \delta_i - \delta_{i-1}$. This notation represents the localization of the extra $r_i$ samples that are trimmed away when the trimming level is increased from $\delta_{i-1}$ to $\delta_i$. 
Finally, let
${\rm \bf y}_i(t) = [y(t-j_1(t)),\ldots,y(t-j_{r_i}(t))]^\T$ denote the $r_i\times(mn)$ vector of such extra samples, and let
$\bPsi_i(t)=[\bpsi(t,j_1(t))|\ldots|\bpsi(t,j_{r_i}(t))]$ represent the $(mn)\times r_i$ matrix of the corresponding generalized regression vectors.

It is straightforward to verify that
\begin{align*}
\P_{{\rm R}|i-1}(t) &= \P_{{\rm R}|i}(t) + \bPsi_i(t)\bPsi_i^\H(t)\\
\p_{{\rm R}|i-1}(t) &= \p_{{\rm R}|i}(t) + \bPsi_i(t){\rm \bf y}_i^*(t)\\
i&=p,\ldots,2
\end{align*}
with initial conditions
\begin{align*}
 \P_{{\rm R}|p}(t) &= \sum_{j\in \Omega_p(t)} \bpsi(t,j) \bpsi^\H(t,j)\\
 \p_{{\rm R}|p}(t) &= \sum_{j\in \Omega_p(t)} y^*(t+j)\bpsi(t,j).
\end{align*}
Using the matrix inversion lemma (Woodbury's identity), one arrives at the following formula for direct recursive computation 
of $\P_{{\rm R}|i-1}^{-1}(t)$
\begin{align}
\begin{split}
\P_{{\rm R}|i-1}^{-1}(t)&=\P_{{\rm R}|i}^{-1}(t)  - \P_{{\rm R}|i}^{-1}(t)\bPsi_i(t)\times\\
&\times [\I_{r_i} + \bPsi_i^\H(t)\bPsi_i(t)]^{-1}
\bPsi_i^\H(t)\P_{{\rm R}|i}^{-1}(t)\\
i&=p,\ldots,2.
\end{split}
\end{align}
While the computational cost of evaluating $\P_{{\rm R}|p}^{-1}(t)$ and $\p_{{\rm R}|p}^{-1}(t)$ is of order ${\mathcal O}(m^3n^3)$ and ${\mathcal O}(m^2n^2)$, respectively,
for $\P_{{\rm R}|i-1}(t)$ and $\p_{{\rm R}|i-1}(t)$, $i=p,\ldots,2$, it reduces to ${\mathcal O}(r_im^2n^2) + {\mathcal O}(r_i^3)$ and ${\mathcal O}(r_imn)$, respectively.
When $r_i \ll mn$, this significantly lowers the computational burden when implementing $p$ trimmed LBF algorithms in a parallel estimation scheme. 

\section{Computer simulations}

Simulations were conducted for a 10-tap FIR model of a self-interference UWA channel. The time-varying impulse 
response coefficients were generated by lowpass filtering zero-mean circular Gaussian noise with exponentially decaying variance 
${\rm var}[\theta_j(t)] = (0.69)^{j-1}$ for $j=1,...,10$, reflecting the power delay profile due to spreading and absorption. 
The bandwidth of the lowpass filter was set to $B=0.003$, corresponding to 3 Hz under 1 kHz sampling (such rates of parameter changes are typical in UWA applications).
The input signal was a white sequence of QPSK symbols, given by $u(t) = (\pm 1 \pm j)/\sqrt{2}$, $\sigma_u^2=1$. 

The trimming level $\delta$ was specified as a percentage of the window width $K$: $\delta= {\rm int}[\mu K]$, where
$\mu=(1-\gamma)$.

In our first identification experiment, the $\epsilon$-contaminated white Gaussian noise was generated according to (\ref{s1a}), with variances $\sigma_1^2=0.032$ and $\sigma_2^2=32$.
Under such settings, in the outlier-free case ($\epsilon=0$) the signal-to noise ratio was equal to ${\rm SNR}=\log_{10} [\sigma_\theta^2/\sigma_1^2]= 20{\rm dB}$.

\begin{figure}
\begin{center}
\includegraphics[scale=0.7]{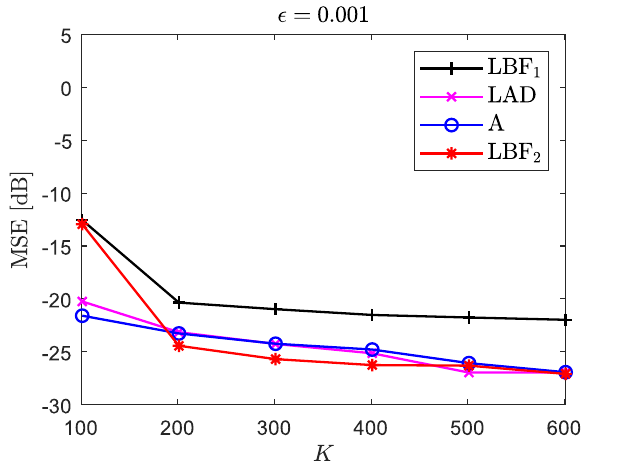}\\[2mm]
\includegraphics[scale=0.7]{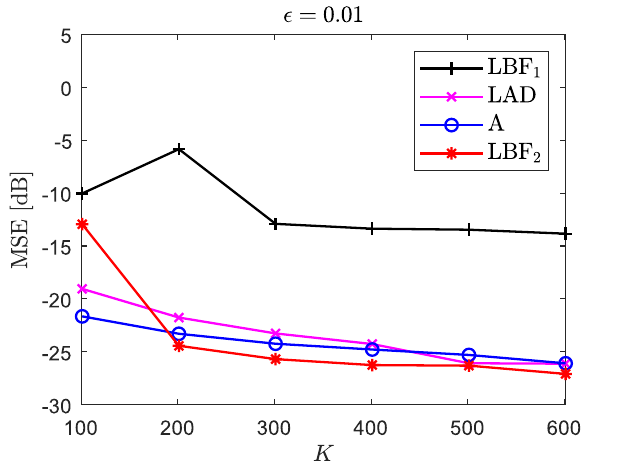}\\[2mm]
\includegraphics[scale=0.7]{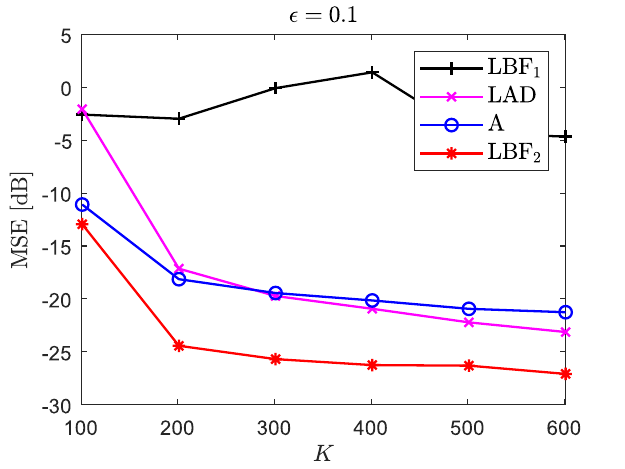}
\end{center}
\vspace{-5mm}
\caption{Comparison of the mean squared parameter tracking errors for three algorithms operating in the presence of $\epsilon$-contaminated
measurement noise: the unmodified LBF algorithm (${\rm LBF}_1$), adaptive trimmed
LBF algorithm (A), and adaptive LAD algorithm. Results obtained for the unmodified LBF algorithm
operating in the absence of impulsive disturbances (${\rm LBF}_2$) are provided as a reference.}
\label{fig1}
\end{figure}

Fig. \ref{fig1} illustrates the dependency of the mean square parameter estimation error (long-time average evaluated over $10^5$ time steps) on $K$. The results are presented for the three identification algorithms discussed in the paper: LBF (in the presence and in the absence of outliers), adaptive trimmed LBF, and adaptive LAD. The plots are shown for $\epsilon=0.001$,
$\epsilon=0.01$ and  $\epsilon=0.1$.
The adaptive trimmed LBF algorithm combined results obtained from 3 trimmed LBF algorithms operating in parallel. These algorithms were tailored for different values of $\mu$, 
specifically $0.5\%$, $5\%$, and $15\%$, respectively. The width of the decision window $L$ was set to 40.
The selection of the number of basis functions was based on (\ref{o10}), using (\ref{103}) and the
trimmed variance estimate (\ref{100}) obtained for $\mu=0.15$ (for trimmed LBF and LAD algorithms)
and $\mu=0$ (for regular LBF algorithms). Fig. 2 presents the influence of the intensity of impulses $\epsilon$ on the MSE for two widths of the analysis window.

Examples of typical noise realizations
and the corresponding parameter estimates are shown in Fig. 3.

Finally, Fig. 4 shows the dependence of the optimal number of basis functions $m_{\rm opt}$ on the width of the
analysis interval $K$ in the outlier-free case ($\epsilon=0$, known variances $\sigma_e^2$ and $\sigma_\theta^2$).

\begin{figure*}[t]
\begin{center}
\includegraphics[scale=0.7]{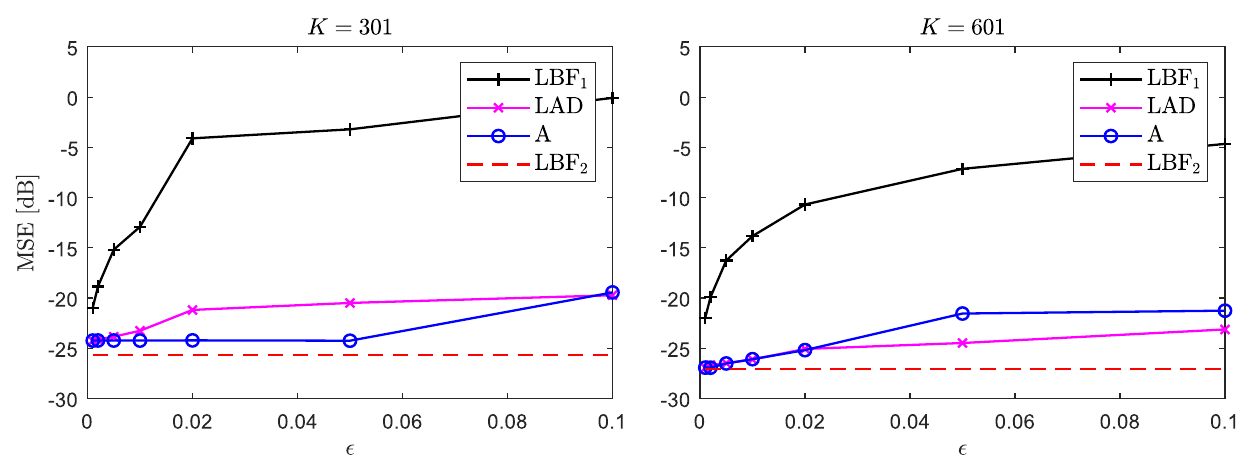}
\end{center}
\vspace{-5mm}
\caption{Comparison of the mean squared parameter tracking errors for three algorithms operating in the presence of $\epsilon$-contaminated
measurement noise: the unmodified LBF algorithm (${\rm LBF}_1$), adaptive trimmed
LBF algorithm (A), and adaptive LAD algorithm. Results obtained for the unmodified LBF algorithm
operating in the absence of impulsive disturbances (${\rm LBF}_2$) are provided as a reference.}
\label{fig2}
\end{figure*}

\begin{figure*}
\begin{center}
\includegraphics[scale=0.75]{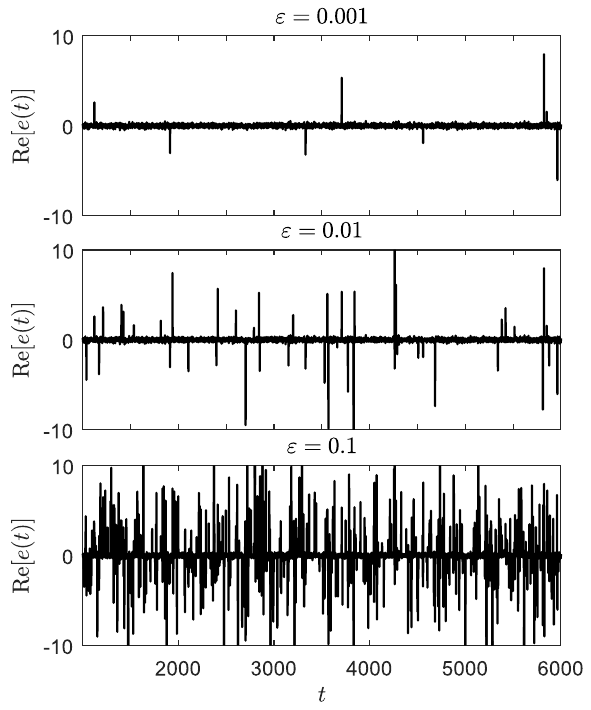} \hspace{20mm}
\includegraphics[scale=0.75]{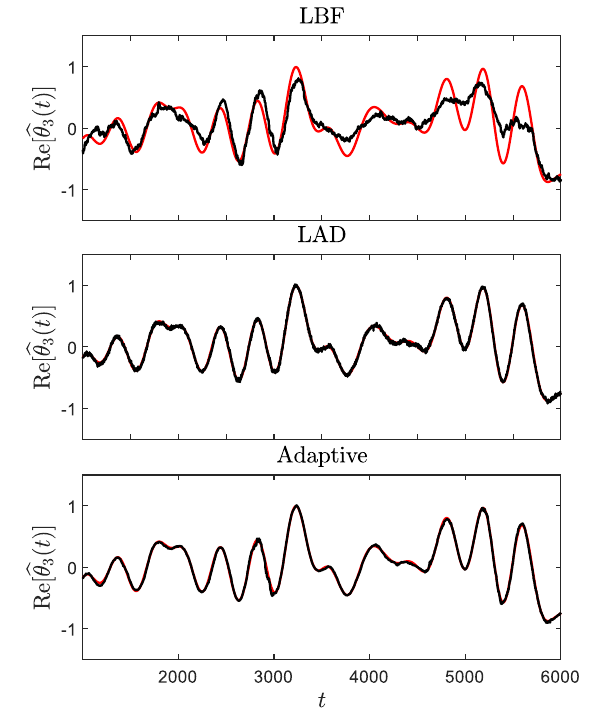}
\vspace{-5mm}
\caption{\hspace{-2mm}Left: typical realizations, displayed in the range [-10, 10], of $\epsilon$-contaminated noise for different values of $\epsilon$. 
Right: estimates (thin red lines) of system parameter ${\rm Re}[\theta_3(t)]$ (thick black lines) obtained for $\epsilon=0.1$ and $K=301$.}
\end{center}
\label{fig3}
\end{figure*}

The adaptive trimmed scheme yields better or only slightly worse results than the computationally much more demanding adaptive LAD scheme. 
Further simulation investigation, not reported here, confirms that the comparison shifts
 decidedly  in favor of the adaptive LBF approach when the contamination level is known, at least approximately, allowing for a narrower working range of the trimming level 
$\delta$. Note that the range of $\delta$ adopted in our simulation experiment was very wide, enabling the effective elimination of outliers  with a wide range of intensities.
Both aforementioned schemes provide significantly better estimation accuracy than the unmodified LBF scheme, denoted by ${\rm LBF}_1$. For comparison purposes, the results yielded by the LBF algorithm operating in the absence of impulsive disturbances ($\epsilon=0$), denoted by ${\rm LBF}_2$, are also shown.

\begin{figure}[h]
\begin{center}
\includegraphics[scale=0.7]{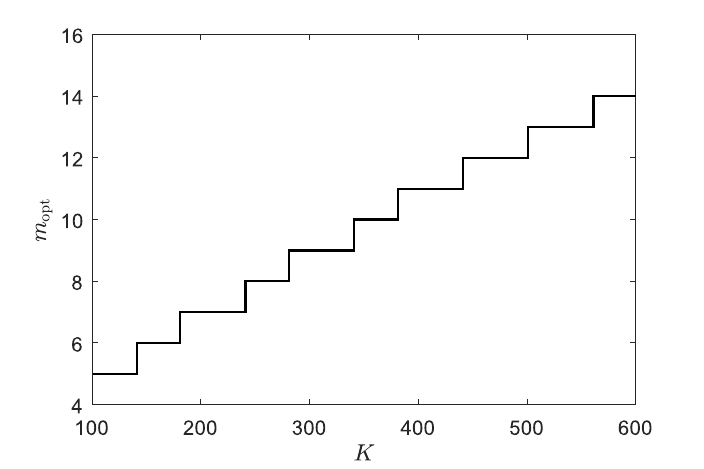}
\end{center}
\vspace{-5mm}
\caption{Dependence of the optimal number of basis functions $m_{\rm opt}$ on the width of the
analysis interval $K$ in the outlier-free case ($\epsilon=0$).}
\label{fig4}
\end{figure}

In the second experiment, the measurement noise followed an $\alpha$-stable distribution.
We note that such a situation may occur, for example, in some communication \cite{ilow}--\cite{hughes} and radar \cite{radar} systems.
More precisely, for all $t$, $e_{\rm R}(t)= {\rm Re}[e(t)]$ and  $e_{\rm I}(t)= {\rm Im}[e(t)]$ were independent random variables with a common
symmetric $\alpha$-stable (S$\alpha$S) distribution $S(\alpha,\sigma)$, where $\alpha$, $0<\alpha \le 2$, is the stability index and $\sigma>0$ is the scale parameter \cite{stable}.
The characteristic function of this distribution has the form
\begin{align*}
\phi(z) ={\rm exp}(-\sigma^\alpha |z|^\alpha).
\end{align*}
The $\alpha$-stable distributions are heavy-tailed and, therefore, well-suited to represent measurement noise containing outliers. They lack closed-form probability and cumulative distribution functions, with the only exceptions being the Gaussian distribution ($\alpha=2$) and the Cauchy distribution ($\alpha=1$).
The degree of ``impulsiveness'' of S$\alpha$S noise increases with decreasing $\alpha$. For $\alpha <2$ the variance of the $\alpha$-stable distribution is infinite,
and for $\alpha \le 1$ its expected value does not exist.
Note that, due to its infinite variance, the $\alpha$-stable noise does not satisfy assumption (A2). 
Experiments with such heavy-tailed noise were conducted to assess the robustness of the proposed 
approach against nonstandard (and more challenging to handle) noise distributions.

Fig. 5 shows the MSE  plots obtained for $\alpha=1.6$, $\alpha=1.4$ and $\alpha=1.2$ (averages evaluated over $10^5$ time steps), and Fig. 6 shows the MSE plots obtained for two widths of analysis window, i.e. $K=301$ and $K=601$ and different values of $\alpha$.
As remarked earlier, the quantile (trimmed) conditional variance remains well-defined in the $\alpha$-stable case \cite{quantile}.
The scale parameter was set to $\sigma=0.09$, such that for $\alpha=2$, which corresponds to the Gaussian (outlier-free) case, the SNR was equal to 20 dB.
For $\alpha<2$, SNR is not well-defined.
Again, the adaptive selection of the number of basis functions was based on (\ref{o10}).

Typical noise realizations and 
the corresponding parameter estimates are shown in Fig. 7.
The conclusions are similar to those drawn from the first experiment.

\begin{figure}[ht]
\begin{center}
\includegraphics[scale=0.7]{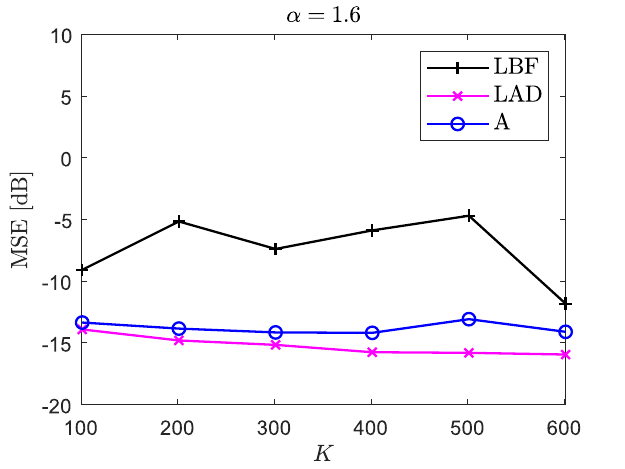}\\[2mm]
\includegraphics[scale=0.7]{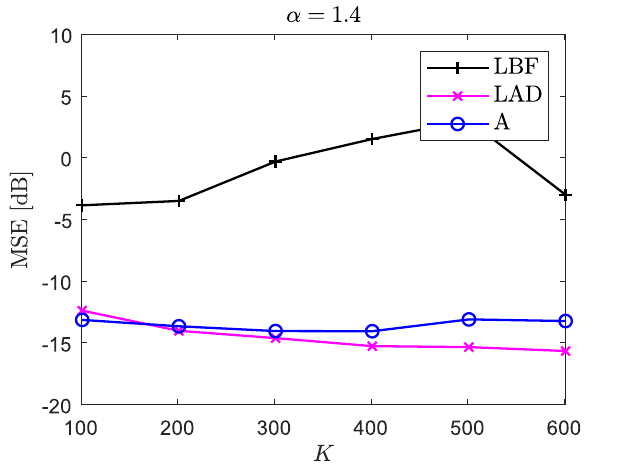}\\[2mm]
\includegraphics[scale=0.7]{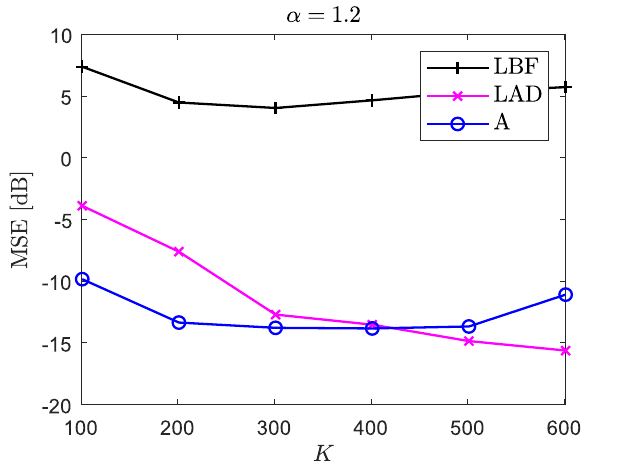}
\end{center}
\vspace{-5mm}
\caption{Comparison of the mean squared parameter tracking errors for three algorithms operating in the presence of $\alpha$-stable
measurement noise: the unmodified LBF algorithm (LBF), adaptive trimmed
LBF algorithm (A), and adaptive LAD algorithm.} 
\label{fig5}
\vspace{-5mm}
\end{figure}

Fig. 8 shows the dependence of the average processing time of a single frame of data on the frame width $K$ for three algorithms: LBF, adaptive trimmed LBF
combining results yielded by three robust LBF algorithms, and adaptive LAD. The results are presented for the $\epsilon$-contaminated disturbance with a contamination level $\epsilon=0.1$,
in the case where the variances $\sigma_e^2$ and $\sigma_\theta^2$ are known. 
The processing times only marginally depend on the type and strength of impulsive noise.
The speed test was run on a computer equipped with an Intel(R) Core(TM) i7-8700 CPU, with a clock speed of 3.2 GHz and 16 GB of RAM.

\begin{figure*}
    \begin{center}
    \includegraphics[scale=0.7]{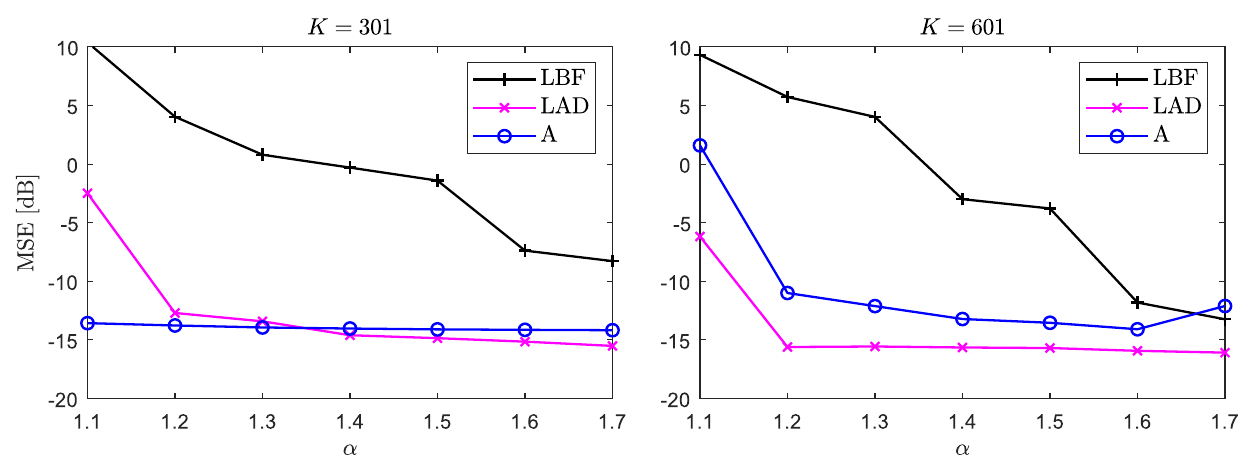}
    \caption{Comparison of the mean squared parameter tracking errors for three algorithms operating in the presence of $\alpha$-stable measurement noise: the unmodified LBF algorithm (LBF), adaptive trimmed
LBF algorithm (A), and adaptive LAD algorithm.}
    \end{center}
    \label{fig6}
\end{figure*}

\begin{figure*}
\vspace{-3mm}
\begin{center}
\includegraphics[scale=0.75]{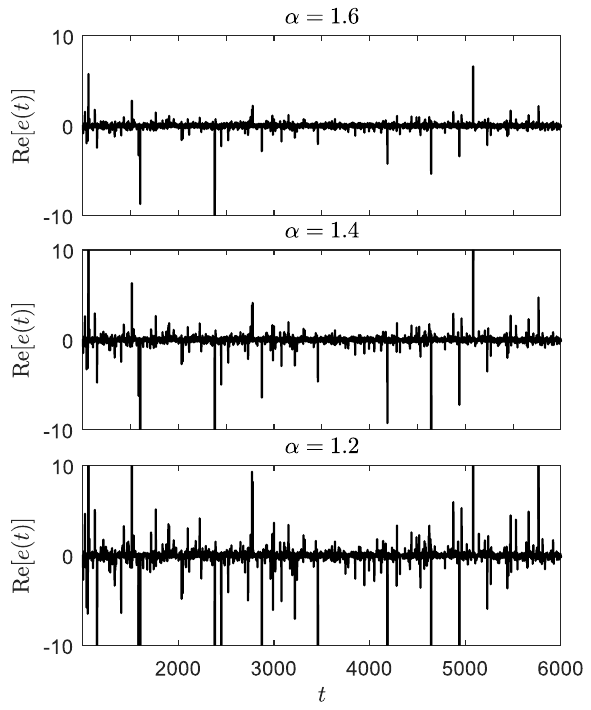} \hspace{20mm}
\includegraphics[scale=0.75]{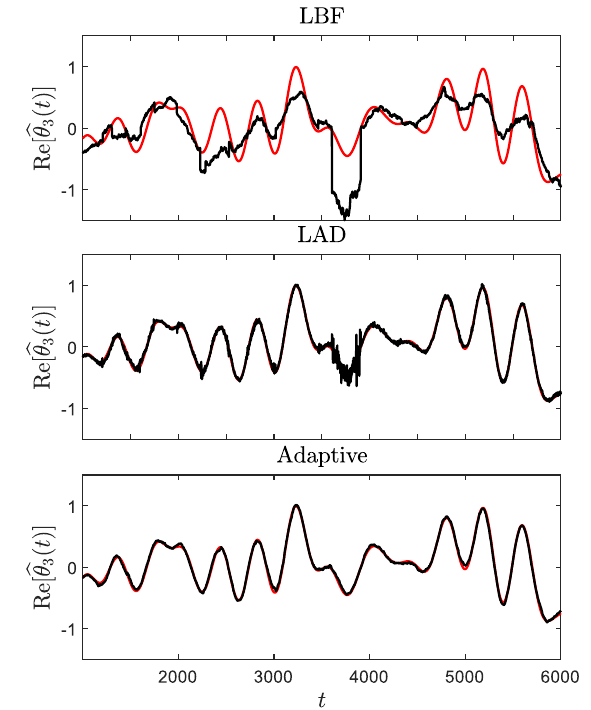}
\vspace{-5mm}
\caption{\hspace{-2mm}Left: typical realizations, displayed in the range [-10, 10], of $\alpha$-stable noise
for three values of $\alpha$. Right: estimates (thin red lines) of system parameter ${\rm Re}[\theta_3(t)]$ (thick black lines) obtained for $\alpha=1.2$ and $K=301$.}
\end{center}
\vspace{-5mm}
\label{fig7}
\end{figure*}

For calculating the LAD estimates, we utilized the Matlab \textit{fmincon} function. At the initial time step, the function was initialized with the LBF estimates of basis function coefficients. Subsequently, in the following time steps, the optimization procedure was initiated with the LAD estimates of basis function coefficients from the previous time instant. For each position of the analysis window, the optimization procedure was stopped after 10 iterations of the optimization algorithm. The average processing time of the LAD algorithm is 
two orders of magnitude greater than that of the adaptive trimmed LBF algorithm.

\section{Conclusion}

The problem of identifying a time-varying FIR system in the presence of impulsive disturbances was addressed and resolved using the local basis function (LBF) approach. First, it was shown that when statistical knowledge about parameter variation is available, the parameter tracking performance of LBF algorithms can be optimized by selecting the most appropriate number and shape of basis functions. The adaptive version of the basis selection rule was then incorporated into two robust, outlier-resistant estimation schemes based on the least squares approach with sequential data trimming and the least absolute deviations (LAD) approach, respectively. It was shown that the sequentially trimmed LBF algorithm provides results that are better than, or comparable to, those yielded by the computationally much more demanding LAD algorithm.

\vskip 5mm
\noindent{\bf Appendix 1}
\vskip 3mm
{\bf \noindent{a)} Derivation of (\ref{o6a})}
\vskip 1mm
Note that 
\begin{align*}
B_{\rm LBF}(m) = \sum_{i=1}^n \E[|\theta_i(t) - \bar{\theta}_i(t)|^2]
\end{align*}
where $[\bar{\theta}_1(t),\ldots,\bar{\theta}_n(t)]^\T= \bar{\btheta}_{\rm LBF}(t)$. According to (\ref{o1})
\begin{align}
&\E[|\theta_i(t) - \bar{\theta}_i(t)|^2] = \sigma_{\theta_i}^2 \Big[ \rho_\theta(0) - \sumk h(j) \rho_\theta(j)\nonumber\\
&- \sumk h^*(j) \rho_\theta^*(j) + \sum_{i=-k}^k \sumk h(i)h^*(j) \rho_\theta(i-j)\Big]\nonumber\\[1mm]
&= \sigma_{\theta_i}^2 [1- \r_\theta^\T\h - \r_\theta^\H \h^* + \h^\H\R_\theta \h],
\label{ap1}
\end{align}
where $\h=[h(k),\ldots,h(-k)]^\T$ and $\r_\theta=[\rho_\theta(k), \ldots,$ $\rho_\theta(-k)]^\T$.
It holds that
\begin{align*}
\R_\theta = \Q\bLambda\Q^\H = \sum_{i=1}^K \lambda_i\q_i\q_i^\H,
\end{align*}
where $\Q=[\q_1|...|\q_K]$ is the $K\times K$ matrix made up of the eigenvectors of $\R_\theta$, and $\lambda_1\ge \ldots \ge \lambda_K >0$ denote
the corresponding eigenvalues.
It is straightforward to verify that in the considered case, $\h=\Q_m\Q_m^\H \ce$, where $\Q_m=[\q_1|...|\q_m]$ is the $K\times m$ matrix composed of the $m$ dominant
eigenvectors of $\R_\theta$, and $\ce=[0,...0,1,0,...]^\T$ is the $K\times m$ vector with only one nonzero element located in its center.
Note that
\begin{align*}
\Q_m\Q_m^\H = \sum_{i=1}^m \q_i\q_i^\H
\end{align*}
resulting in
\begin{align}
\h^\H\R_\theta\h &= \ce^\T \left(\sum_{i=1}^m \q_i\q_i^\H \right)\left(\sum_{i=1}^K \lambda_i \q_i\q_i^\H\right)
\left(\sum_{i=1}^m \q_i\q_i^\H \right)\ce \nonumber\\
&=\sum_{i=1}^m \lambda_i \ce^\T\q_i\q_i^\H\ce,
\label{ap2}
\end{align}
where the second transition follows from the fact the eigenvectors of $\R_\theta$ are orthonormal, i.e., $\q_i^\H\q_j=0$ for $i\neq j$
and $\q_i^\H\q_i=1$. Finally, note that since  $\r_\theta= \R_\theta^*\ce$, it holds that
\begin{align}
\r_\theta^\T\h &= \ce^\T\R_\theta\Q_m\Q_m^\H\ce = \ce^\T \left(\sum_{i=1}^K \lambda_i \q_i\q_i^\H\right)
\left(\sum_{i=1}^m \q_i\q_i^\H \right)\ce\nonumber\\
&= \sum_{i=1}^m \lambda_i \ce^\T\q_i\q_i^\H\ce = \r_\theta^\H\h^*.
\label{ap3}
\end{align}
Combining (\ref{ap1}), (\ref{ap2}) and (\ref{ap3}), and noting that $\ce^\T\q_i=f_i(0)$,
one arrives at (\ref{o6a}).

\vskip 3mm
{\bf \noindent{b)} Derivation of (\ref{o7a})}

Note that 
\begin{align}
\sumk |h(j)|^2 &= \f^\H(0) \left( \sumk \f(j)\f^\H(j)\right)\f(0)\nonumber\\
&= \f^\H(0)\f(0) = \sum_{i=1}^m f_i^2(0).
\label{ap4}
\end{align}
Hence, combining (\ref{o2}) and (\ref{ap4}), one obtains (\ref{o7a}).

\vskip 5mm
\noindent{\bf Appendix 2}
\vskip 3mm
\noindent{Denote} $\widehat{\btheta}_{\rm R}(t+j|t)= [\widehat{\theta}_1(t+j|t), \ldots, \widehat{\theta}_n(t+j|t)]^\T$,
$\widehat{\bbeta}_{\rm R}(t)= [\widehat{\bbeta}_1^\T(t), \ldots, \widehat{\bbeta}_n^\T(t)]^\T$ and
$\bar{\btheta}_{\rm R}(t)=[\bar{\theta}_1(t),\ldots,\bar{\theta}_n(t)]^\T$.
It is straightforward to check that
\begin{align}
\widehat{\sigma}_\theta^2(t) &= \frac{1}{K} \sumk ||\widehat{\btheta}_{\rm R}(t+j|t) - \bar{\btheta}_{\rm R}(t)||^2\nonumber\\
 &= \frac{1}{K} \sumk ||\widehat{\btheta}_{\rm R}(t+j|t)||^2  - ||\bar{\btheta}_{\rm R}(t)||^2
\label{b1}
\end{align}
Since $\widehat{\theta}_i(t+j|t)= \f^\H(j)\widehat{\bbeta}_i(t)$, exploiting (\ref{3}), one obtains
\begin{align*}
\sumk |\widehat{\theta}_i(t+j|t)|^2 &= \widehat{\bbeta}_i^\H(t) \left( \sum_{j=-k}^k \f(j)\f^\H(j)\right)\widehat{\bbeta}_i(t) \nonumber\\
&= ||\widehat{\bbeta}_i(t)||^2
\end{align*} 
leading to
\begin{align}
\sumk ||\widehat{\btheta}_{\rm R}(t+j|t)||^2 = \sum_{i=1}^n ||\widehat{\bbeta}_i(t)||^2 = ||\widehat{\bbeta}_{\rm R}(t)||^2.
\label{b2}
\end{align}
To compute the second component on the right-hand side of (\ref{b1}), note that
\begin{align*}
\bar{\theta}_i(t) = \frac{1}{K} \sumk \widehat{\theta}_i(t+j|t)= \g^\H \widehat{\bbeta}_i(t)
\end{align*}
where $\g= (1/K) \sumk \f(j)$. This results in
\begin{align}
||\bar{\btheta}_{\rm R}(t)||^2 = \sum_{i=1}^n |\bar{\theta}_i(t)|^2 = || \G \widehat{\bbeta}_{\rm R}(t)||^2
\label{b3}
\end{align}
where $\G=\I_n\otimes \g^\H$. Combining (\ref{b1}), (\ref{b2}), and (\ref{b3}), one arrives at (\ref{103}).

\begin{figure}[t]
\begin{center}
\includegraphics[scale=0.7]{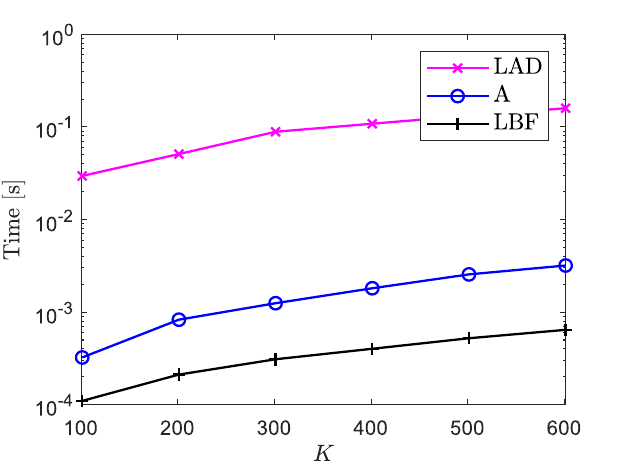}
\end{center}
\vspace{-5mm}
\caption{Dependence of the average processing time of a single frame of data on the frame width $K$ for three algorithms: LBF, adaptive LAD, and adaptive trimmed LBF (A).}
\label{fig8}
\end{figure}

\begin{IEEEbiography}[{\includegraphics[width=1in,height=1.25in,clip,keepaspectratio]{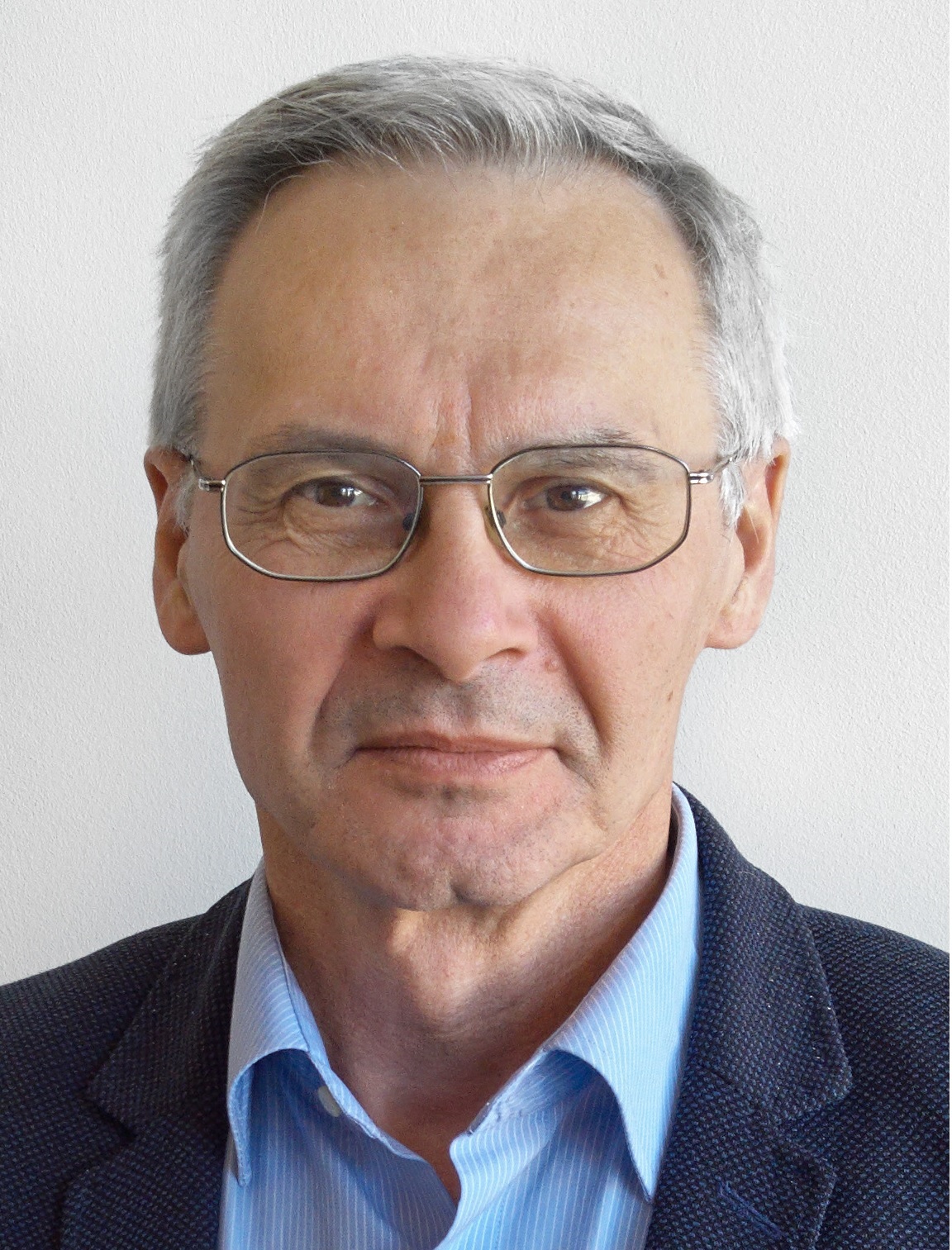}}]{Maciej Nied\'zwiecki}
(M'08, SM'13)
received the M.Sc. and Ph.D. degrees from the Technical University of Gda\'nsk,
Gda\'nsk, Poland and the Dr.Hab. (D.Sc.) degree from the Technical University 
of Warsaw, Warsaw, Poland, in 1977, 1981 and 1991, respectively.
He spent three years as a Research Fellow with the Department of 
Systems Engineering, Australian National University, 1986-1989.
In 1990 - 1993 he served as a Vice Chairman of Technical Committee on Theory
of the International Federation of Automatic Control (IFAC). 
He is the author of the book {\it Identification of
Time-varying Processes} (Wiley, 2000).  His main areas of 
research interests include system identification, statistical signal processing and 
adaptive systems.

Dr. Nied\'zwiecki is currently a member of the
IFAC committees on Modeling, Identification and Signal Processing and on Large 
Scale Complex Systems, and 
a member of the Automatic Control and Robotics Committee of the 
Polish Academy of Sciences (PAN).
He works as a Professor at the Department of Signals and Systems, 
Faculty of Electronics, Telecommunications and Informatics, 
Gda\'nsk University of Technology.
\end{IEEEbiography}

\begin{IEEEbiography}[{\includegraphics[width=1in,height=1.25in,clip,keepaspectratio]{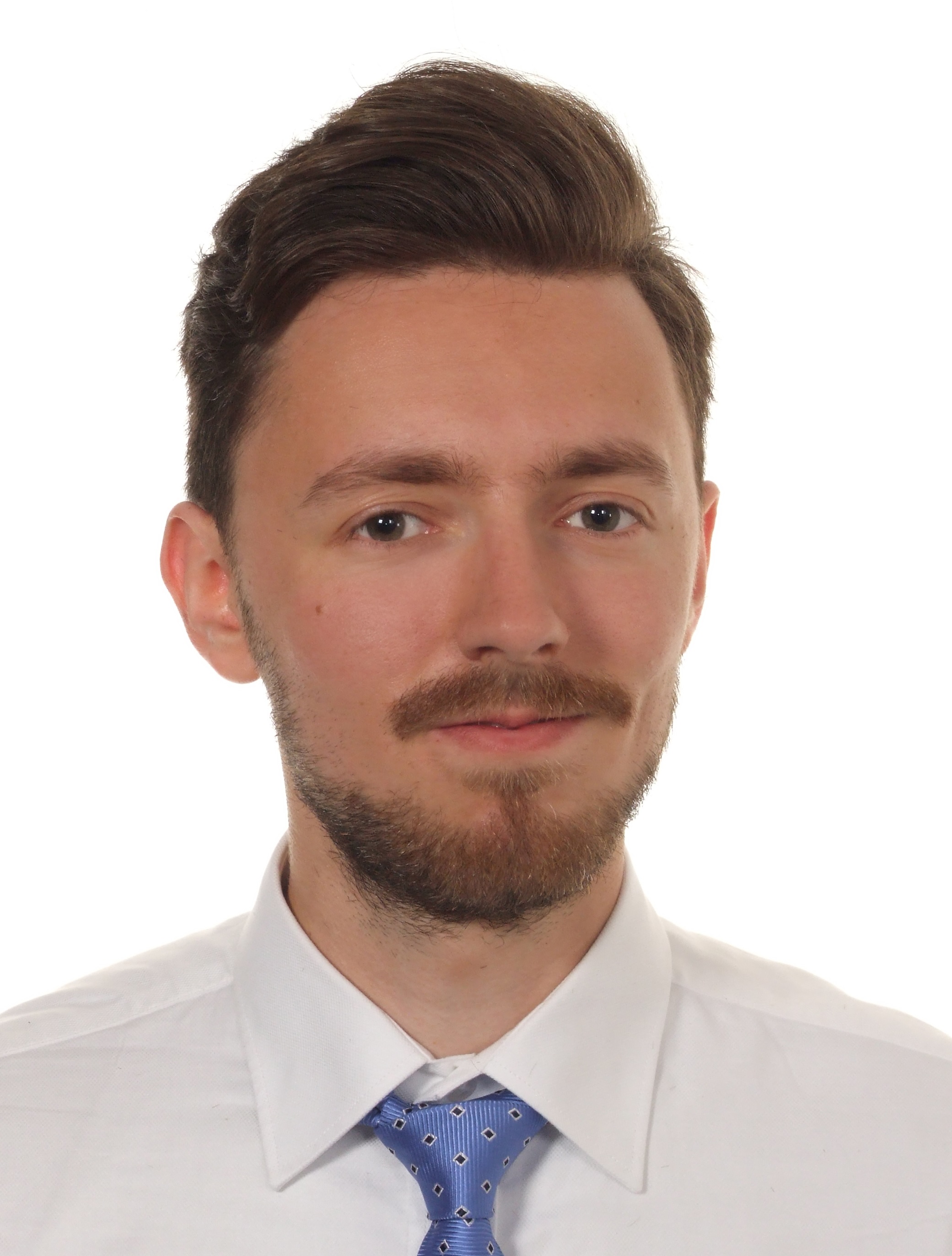}}]{Artur Ga\'ncza }
(M'23)
earned his M.Sc. and Ph.D. degrees
from the Gda\'nsk University of
Technology (GUT), Gda\'nsk, Poland, in 2019 and 2024, respectively.
He currently serves as an Assistant Professor in the Department of Signals and Systems at the Faculty of Electronics, Telecommunications, and Informatics.
His research interests include optimization, system identification, and adaptive signal processing.
\end{IEEEbiography}

\begin{IEEEbiography}[{\includegraphics[width=1in,height=1.25in,clip,keepaspectratio]{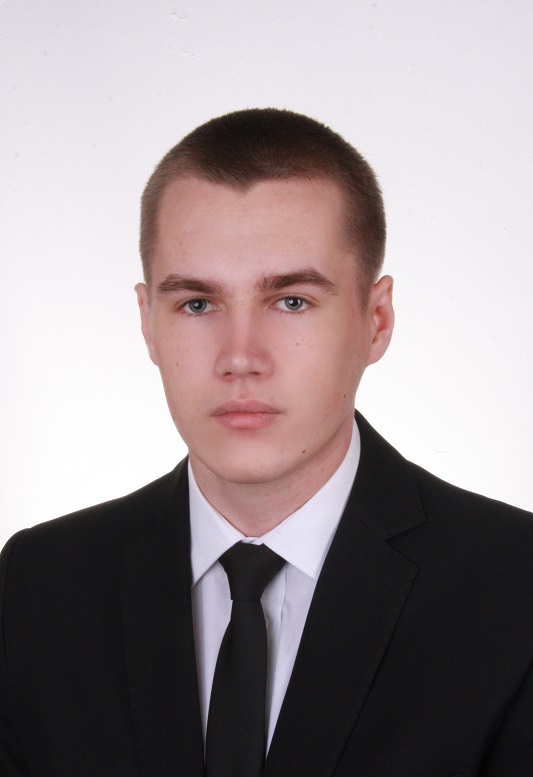}}]{Wojciech \.Zu\l awi\'nski }
 earned Ph.D. degree in mathematics at Wroclaw University of Science and Technology (WroclawTech) in 2025. He holds master’s degrees in applied mathematics and big data 
 analytics from WroclawTech. His research interests include time series analysis, heavy-tailed distributions, and robust statistics. Primarily, his research 
 focuses on non-Gaussian cyclostationary processes and their applications in real data analysis.
\end{IEEEbiography}

\begin{IEEEbiography}[{\includegraphics[width=1in,height=1.25in,clip,keepaspectratio]{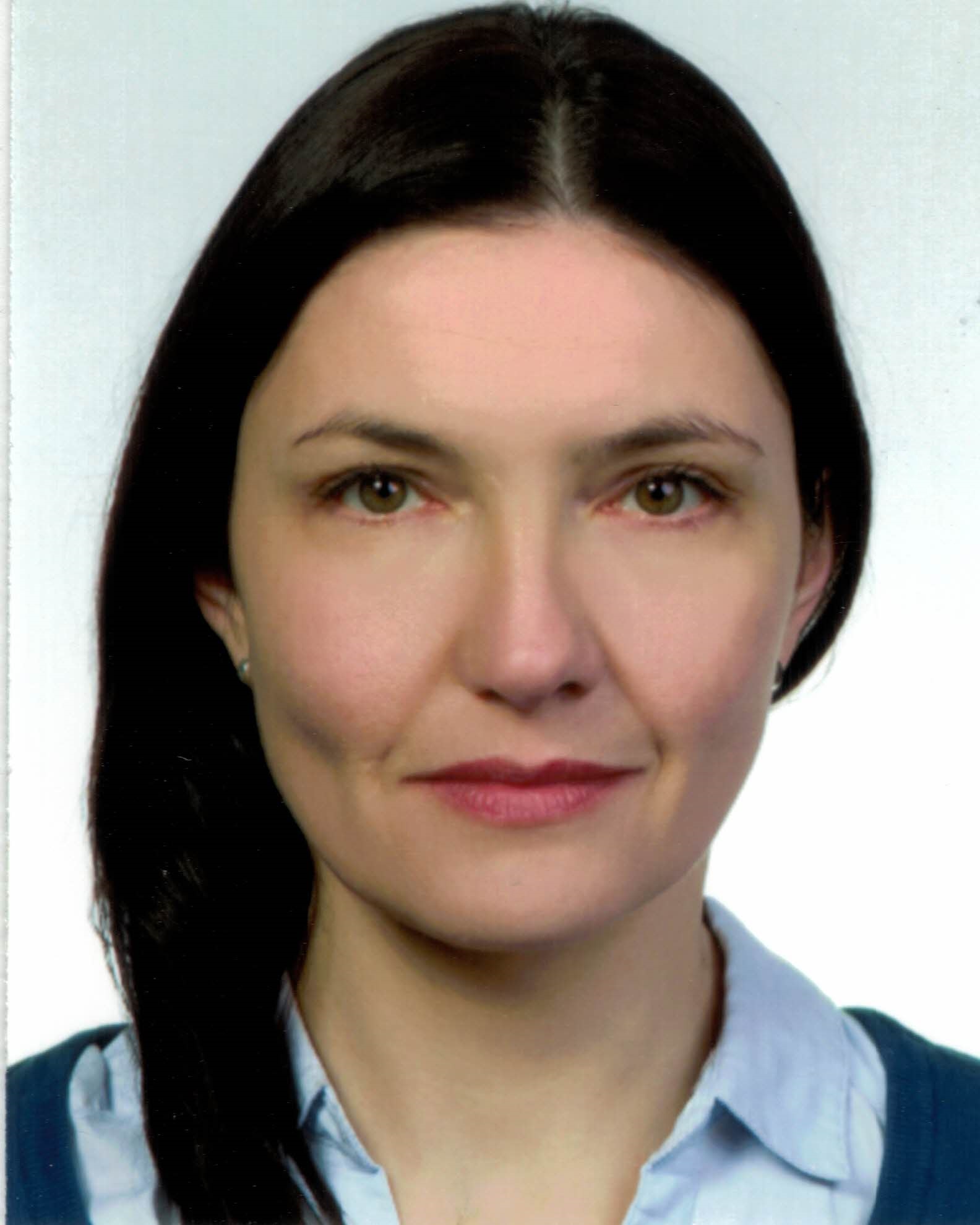}}]{Agnieszka Wy\l oma\'nska }
 received the Ph.D. degree in mathematics and the D.Sc. degree in mining and geology from Wroclaw University of Science and Technology (WroclawTech), in 2006 and in 2015, respectively. 
 She is currently
a Professor with the Faculty of Pure and Applied Mathematics, WroclawTech, and a member of the Hugo Steinhaus Center for Stochastic Processes. She is the author of more than 
200 research papers (according to Scopus) in the area of applied and
industrial mathematics. Her research interests include time-series analysis, stochastic modeling, and statistical analysis of real data with heavy-tailed behavior. 
She is the Editor-in-Chief of Mathematica Applicanda.
\end{IEEEbiography}


\vspace{-2mm}

\begin{thebibliography}{99}

\bibitem{tsatsa} M. K. Tsatsanis and G. B. Giannakis, ``Modelling and equalization of rapidly fading channels,'' 
{\it Int. J. Adapt. Contr. Signal Process.}, vol. 310,
pp. 515-522, 1996. 

\bibitem{proakis} 
J.G. Proakis, {\em Digital communications} (4th ed.), McGraw-Hill, 2001.

\bibitem{siderius}
M.~Siderius and M.~B. Porter, ``Modeling broadband ocean acoustic transmissions
with time-varying sea surfaces,'' \emph{J. Acoust. Soc. Amer.}, vol. 124, no.~1, pp. 137--150, 2008.

\bibitem{uwa} 
M. Stojanovic and J. Preisig, ``Underwater acoustic communication channels: Propagation models 
and statistical characterization,'' {\it IEEE Comm. Mag.}, vol. 47, pp. 84-89, 2009.

\bibitem{book}
M.~{Nied\'zwiecki}, \emph{{Identification of Time-varying Processes}}.\hskip
  1em plus 0.5em minus 0.4em\relax {New York}: Wiley, 2000.

\bibitem{dahlhaus2012}
R. Dahlhaus, ``Locally stationary processes,'' {\em Handbook Statist.}, vol. 25,
pp. 1-37, 2012.

\bibitem{haykin}
S. Haykin, {\em Adaptive Filter Theory}. Prentice-Hall, 1996.

\bibitem{sayed}
A. H. Sayed,
{\em Fundamentals of Adaptive Filtering},
Wiley, 2003.


\bibitem{rao}
T. Subba Rao, ``The fitting of nonstationary time-series models with
time-dependent parameters,'' \textit{J. R. Statist. Soc. B}, vol. 32, pp. 312–322,
1970.

\bibitem{mendel}
J. M. Mendel, \textit{Discrete Techniques of Parameter Estimation: The Equation
Error Formulation}, New York: Marcel Dekker, 1973.

\bibitem{liporace}
L. A. Liporace, ``Linear estimation of nonstationary signals,'' \textit{J. Acoust.
Soc. Amer.}, vol. 58, pp. 1288–1295, 1975.


\bibitem{charbonnier}
R. Charbonnier, M. Barlaud, G. Alengrin and J. Menez,
``Results on AR-modelling of non-stationary signals,''
\textit{Signal Process.}, vol. 12, pp. 143-151, 1987.

\bibitem{functional}
M. Nied\'zwiecki, ``Functional series modeling approach to identification of nonstationary
stochastic systems,'' \textit{IEEE Trans. Automat. Contr.}, vol. 33, pp.
955–961, 1988.

\bibitem{recursiveBF}
M. Nied\'zwiecki, ``Recursive functional series modeling estimators for identification of time-varying plants -- 
more bad news than good?,'' \textit{IEEE Trans. Automat. Contr.}, vol. 35, pp.
610–-616, 1990.

\bibitem{dahlhausbf}
R. Dahlhaus, ``Fitting time series models to nonstationary processes,'' 
{\em Ann. Statist.}, vol. 25,
pp. 1-37, 1997.

\bibitem{mrad}
R. B. Mrad, S.D. Fassois, and J.A. Levitt,
``A polynomial-algebraic method for non-stationary TARMA signal analysis–Part I: The method,''
\textit{Signal Process.}, vol. 65, pp. 1--19, 1998.

\bibitem{J58}
M. Nied\'zwiecki and S. Gackowski,
``New approach to noncausal identification of nonstationary FIR systems subject to
both smooth and abrupt parameter changes,'' 
{\em IEEE Trans. Automat. Contr.}, vol. 58, pp. 1847-1853, 2013. 

\bibitem{norton} 
J. P. Norton, ``Optimal smoothing in the identification of linear time-varying systems,''
{\em Proc. IEE}, vol. 122, pp. 663–668, 1975.

\bibitem{kitagawa}
G. Kitagawa and W. Gersch, ``A smoothness priors time-varying AR
coefficient modeling of nonstationary covariance time series,''
{\em IEEE Trans. Automat. Contr.}, vol. 30, pp. 48–56, 1985.

\bibitem{niedz5}
M. Nied\'zwiecki, ``Locally adaptive cooperative Kalman smoothing and its 
application to identification of nonstationary stochastic systems,'' {\em IEEE Trans. Signal
Process.}, vol. 60, pp. 48-59, 2012.

\bibitem{genSG} M. Nied\'zwiecki and M. Cio\l ek, ``Generalized  Savitzky-Golay  filters for identification of nonstationary systems,'' 
{\it Automatica}, vol. 108, No. 108477, 2019.

\bibitem{mechanical}
W. \.Zu\l awi\'nski, J. Antoni, R. Zimroz, and A. Wyłoma\'nska, 
``Robust coherent and incoherent statistics for detection of hidden periodicity in models with non-Gaussian additive noise,'' \textit{EURASIP Journal on Advanced Signal Processing}, 
No. 71, 2024.


\bibitem{huber} P.J. Huber and E. Ronchetti,  \textit{Robust Statistics}, (2nd ed.), Wiley, 2009.

\bibitem{lts} P.J. Rousseeuw and A.M. Leroy, \textit{Robust Regression and Outlier Detection}, Wiley, 2005.

\bibitem{zou}
Y. Zou, S. Chan, and T. Ng, ``A recursive least M-estimate (RLM) adap-
tive filter for robust filtering in impulse noise,'' IEEE Signal Processing
Letters, vol. 7, no. 11, pp. 324–326, 2000


\bibitem{chan}
S.-C. Chan and Y.-X. Zou, ``A recursive least M-estimate algorithm for
robust adaptive filtering in impulsive noise: fast algorithm and conver-
gence performance analysis,'' IEEE Transactions on Signal Processing,
vol. 52, no. 4, pp. 975–991, 2004


\bibitem{combination}
Á. Navia-Vazquez and J. Arenas-Garcia, ``Combination of recursive least
p-norm algorithms for robust adaptive filtering in alpha-stable noise,''
IEEE Trans. Signal Process., vol. 60, no. 3, pp. 1478–1482, 2012.


\bibitem{hur}
J. Hur, I. Song, and P. Park, ``A variable step-size normalized subband
adaptive filter with a step-size scaler against impulsive measurement
noise,'' IEEE Trans. Circuits Syst. II, Exp. Briefs, vol. 64, no. 7,
pp. 842–846, 2017.

\bibitem{yu}
Y. Yu, Z. Zheng, W. Wang, Y. Zakharov, and R.S. de Lamare,
``DCD-based recursive adaptive algorithms robust against impulsive noise,''
IEEE Trans. Circuits Syst. II, Exp. Briefs, vol. 67, no. 7,
pp. 1359–1363, Jul. 2020.


\bibitem{schafer}
R.W. Schafer, ``What is a Savitzky-Golay filter?,'' \textit{IEEE Signal Process. Mag.}, vol. 28, pp. 111-117, 2011.


\bibitem{shen}
L.~Shen, B.~Henson, Y.~Zakharov, and P.~Mitchell, ``Digital self-interference
cancellation for full-duplex underwater acoustic systems,'' \emph{IEEE
Trans. Circuits Syst. II: Express Briefs}, vol.~67, no.~1, pp.
192-196, 2019. 

\bibitem{uwa1}
L. Shen, Y. Zakharov, B. Henson, N. Morozs and P. Mitchell, ``Adaptive filtering for full-duplex UWA systems with time-varying self-interference channel,'' 
\emph{IEEE Access}, vol. 8, pp. 187590-187604, 2020.

\bibitem{sparse}
M. Nied\'zwiecki, A. Ga\'ncza, L. Shen, L. and Y. Zakharov, ``Adaptive identification of sparse underwater acoustic channels with a
mix of static and time-varying parameters,''
{\it Signal Process.}, vol. 200, p. 108664, 2022.


\bibitem{borah}
D.K. Borah, B.D. Hart, Frequency-selective fading channel estimation
with a polynomial time-varying channel model, {\it IEEE Trans. Comm.},
vol. 47, pp. 862-873, 1999.

\bibitem{wav_tsatsanis} M. K. Tsatsanis and G. B. Giannakis ``Time-varying system identification and model reduction using wavelets,`` 
{\it IEEE Trans. Signal Process.}, vol. 41, pp. 3512-3523, 1994.


\bibitem{billings_2002} H. L. Wei, J. J. Liu and S. A. Billings, ``Identification  of  time-varying systems using multi-resolution wavelet models,'' 
{\it Int. J. Syst. Sci.}, vol. 33, pp. 1217-1228, 2002.



\bibitem{thesis}
A. Ga\'ncza, ``{Local basis function method for identification of nonstationary systems},''
Ph.D. Thesis, Gdansk University of Technology, 2024 (English).


\bibitem{jakes}
W.C. Jakes, Ed., \emph{Microwave Mobile Communication}, Wiley, 1974.


\bibitem{KL}
M. Nied\'zwiecki and A. Ga\'ncza, ``Karhunen-Lo\`eve-based approach to tracking of rapidly fading wireless communication channels,''
{\it Signal Processing}, vol. 209, p. 109043, 2023.

\bibitem{tsp24}
M. Nied\'zwiecki and A. Ga\'ncza, ``On optimal tracking of rapidly varying telecommunication channels,''
{\it IEEE Trans. Signal Process.}, vol. 72, pp. 2726-2738, 2024.

\bibitem{bovik}
A. Restrepo and A. Bovik, ``Adaptive trimmed mean filters for image restoration,''
{\em IEEE Trans. Acoust. Speech Signal Process.}, vol. 36, pp. 1326-1337, 1988.


\bibitem{lts1}
P. \v C{\'i}\v zek and J.A. V{\'i}\v sek, ``Least Trimmed Squares'', pp. 49-62, in 
\textit{XploRe® --- Application Guide}, Springer, 2000.

\bibitem{quantile}
M. Pitera, A. Chechkin, and  A. Wyłoma\'nska,
``Goodness-of-fit test for $\alpha$-stable distribution based
on the quantile conditional variance statistics,'' 
Statistical Methods \& Applications, vol. 31, pp. 387–424, 2022.


\bibitem{lanczos1}
C. Lanczos, \textit{Applied Analysis}, Dover Publications, 1988,
Chap. 2

\bibitem{lanczos2}
G. H. Golub and C. F. VanLoan, \textit{Matrix Computations}, 4th edition, Balti-
The Johns Hopkins University Press, 2013.

\bibitem{trench}
W.F. Trench, ``Numerical solution of the eigenvalue problem for Hermitian Toeplitz matrices,''
SIAM J. Matrix Anal. Appl. vol. 10, pp. 135-156, 1989. 



\bibitem{stable}
G. Samorodnitsky and M.S. Taqqu, {\em Stable Non-{G}aussian Random Processes: Stochastic Models with Infinite Variance}. Chapman and Hall, 1994.


\bibitem{ilow} J. How and D. Hatzinakos, "Analytic alpha-stable noise modeling
in a Poisson field of interferers or scatterers," \textit{IEEE Trans. Signal
Process}, vol. 46, pp. 1601-1611, 1998.

\bibitem{hughes} B. L. Hughes, ``Alpha-stable models of multiuser interference,'' in IEEE International Symposium on Information Theory, Sorrento, Italy,
2000.

\bibitem{radar} R. Raspanti, P. Tsakalides, C.L. Nikias and E. Del Re,
``Cramer-Rao bounds for target angle and Doppler estimation for airborne radar in Cauchy interference,''
Proc. 8th Workshop on Statistical Signal and Array Processing, Corfu, Greece, pp. 534-537, 1996.

\bibitem{CDC24}
M. Nied\'zwiecki, A. Ga\'ncza, W. \.Zu\l awi\'nski and A. Wy\l oma\'nska,
``Robust local basis function algorithms for identification of time-varying FIR systems in impulsive noise environments,''
to  be presented at the 63rd Conference on Decision and Control, December, 16-19, 2024, Milano, Italy. 

\end{thebibliography}
\end{document}